\documentclass[letterpaper,twocolumn,10pt]{article}
\usepackage{usenix2019_v3}

\usepackage{tikz}
\usetikzlibrary{plotmarks}
\usepackage{amsmath,amssymb}

\usepackage[english]{babel}
\usepackage{graphicx,subfigure}
\usepackage{graphics}
\usepackage{soul}
\usepackage{amssymb}
\usepackage{pifont}

\usepackage{amsmath}
\usepackage[cmintegrals]{newtxmath}
\usepackage{url}
\usepackage{listings}
\usepackage[square, numbers]{natbib}
\usepackage{xcolor}
\usepackage{xspace}
\usepackage{booktabs}
\usepackage{tablefootnote}
\usepackage{algorithm}
\usepackage{algorithmic}
\usepackage{adjustbox}
\usepackage{comment}

\usepackage{colortbl}
\usepackage{xintexpr, xinttools}

\newcommand{\para}[1]{\smallskip \noindent \textbf{#1}}
\newcommand{\parait}[1]{\smallskip \noindent \textit{#1}}

\newcommand*\fullcirc[1][1ex]{\tikz\fill (0,0) circle (#1);} 

\newcommand\name{\textsc{WebGraph}\xspace}
\newcommand\adgraph{\textsc{AdGraph}\xspace}

\expandafter\def\expandafter\UrlBreaks\expandafter{\UrlBreaks
  \do\a\do\b\do\c\do\d\do\e\do\f\do\g\do\h\do\i\do\j%
  \do\k\do\l\do\m\do\n\do\o\do\p\do\q\do\r\do\s\do\t%
  \do\u\do\v\do\w\do\x\do\y\do\z\do\A\do\B\do\C\do\D%
  \do\E\do\F\do\G\do\H\do\I\do\J\do\K\do\L\do\M\do\N%
  \do\O\do\P\do\Q\do\R\do\S\do\T\do\U\do\V\do\W\do\X%
  \do\Y\do\Z}

\begin{document}
\date{}

\title{\name: \\Capturing Advertising and Tracking Information Flows for Robust Blocking}

\author{
{\rm Sandra Siby}\\
EPFL
\and
{\rm Umar Iqbal}\\
The University of Iowa
 \and
{\rm Steven Englehardt}\\ 
DuckDuckGo
\and
{\rm Zubair Shafiq}\\
University of California, Davis
\and
{\rm Carmela Troncoso}\\
EPFL
} 

\maketitle

\def\robgraphs{100\xspace}
\def\robgrowth{20\%\xspace}
\def\robbins{5\xspace}
\def\robsamples{20\xspace}
\def\robothers{6\xspace}
\def\robtotdata{540\xspace}
\def\robreqmean{5.98\xspace}
\def\robreqstd{5.39\xspace}
\def\robreqmax{26\xspace}

\def\supadvgraphs{100\xspace}
\def\lonesuccessmean{41.41\xspace}
\def\lonesuccessstd{37.06\xspace}
\def\lonecdmean{0.67\xspace}
\def\lonecdstd{3.31\xspace}
\def\colsuccessmean{33.83\xspace}
\def\colsuccessstd{33.31\xspace}
\def\colcdmean{0.66\xspace}
\def\colcdstd{3.15\xspace}

\definecolor{lightgray}{rgb}{0.95, 0.95, 0.95}
\definecolor{darkgray}{rgb}{0.4, 0.4, 0.4}
\definecolor{editorGray}{rgb}{0.95, 0.95, 0.95}
\definecolor{editorOcher}{rgb}{1, 0.5, 0} 
\definecolor{editorGreen}{rgb}{0, 0.5, 0} 
\definecolor{orange}{rgb}{1,0.45,0.13}		
\definecolor{olive}{rgb}{0.17,0.59,0.20}
\definecolor{brown}{rgb}{0.69,0.31,0.31}
\definecolor{purple}{rgb}{0.38,0.18,0.81}
\definecolor{lightblue}{rgb}{0.1,0.57,0.7}
\definecolor{lightred}{rgb}{1,0.4,0.5}

\lstdefinelanguage{JavaScript}{
  morekeywords={typeof, new, true, false, catch, function, return, null, catch, switch, var, if, in, while, do, else, case, break},
  morecomment=[s]{/*}{*/},
  morecomment=[l]//,
  morestring=[b]",
  morestring=[b]'
}

\lstdefinelanguage{HTML5}{
  language=html,
  sensitive=true,	
  alsoletter={<>=-},	
  morecomment=[s]{<!-}{-->},
  tag=[s],
  otherkeywords={
  >,
	<!DOCTYPE,
  </html, <html, <head, <title, </title, <style, </style, <link, </head, <meta, />,
	</body, <body,
	</div, <div, </div>, 
	</p, <p, </p>,
	</script, <script,
  <canvas, /canvas>, <svg, <rect, <animateTransform, </rect>, </svg>, <video, <source, <iframe, </iframe>, </video>, <image, </image>, <header, </header, <article, </article
  },
  ndkeywords={
  =,
  charset=, src=, id=, width=, height=, style=, type=, rel=, href=,
  fill=, attributeName=, begin=, dur=, from=, to=, poster=, controls=, x=, y=, repeatCount=, xlink:href=,
  margin:, padding:, background-image:, border:, top:, left:, position:, width:, height:, margin-top:, margin-bottom:, font-size:, line-height:,
  transform:, -moz-transform:, -webkit-transform:,
  animation:, -webkit-animation:,
  transition:,  transition-duration:, transition-property:, transition-timing-function:,
  }
}

\lstdefinestyle{htmlcssjs} {%
  basicstyle={\footnotesize\ttfamily},   
  frame=b,
  identifierstyle=\color{black},
  keywordstyle=\color{blue}\bfseries,
  ndkeywordstyle=\color{editorGreen}\bfseries,
  stringstyle=\color{editorOcher}\ttfamily,
  commentstyle=\color{brown}\ttfamily,
  language=HTML5,
  alsolanguage=JavaScript,
  alsodigit={.:;},	
  tabsize=2,
  showtabs=false,
  showspaces=false,
  showstringspaces=false,
  extendedchars=true,
  breaklines=true,
  literate=%
  {Ö}{{\"O}}1
  {Ä}{{\"A}}1
  {Ü}{{\"U}}1
  {ß}{{\ss}}1
  {ü}{{\"u}}1
  {ä}{{\"a}}1
  {ö}{{\"o}}1
}

\begin{abstract}

Millions of web users directly depend on ad and tracker blocking tools to protect their privacy. 
However, existing ad and tracker blockers fall short because of their reliance on trivially susceptible advertising and tracking content.
In this paper, we first demonstrate that the state-of-the-art machine learning based ad and tracker blockers, such as \adgraph, are susceptible to adversarial evasions deployed in real-world. 
Second, we introduce \name, the first graph-based machine learning blocker that detects ads and trackers based on their \textit{action} rather than their content.
By building features around the actions that are fundamental to advertising and tracking -- storing an identifier in the browser, or sharing an identifier with another tracker -- \name performs nearly as well as prior approaches, but is significantly more robust to adversarial evasions.
In particular, we show that \name achieves comparable accuracy to \adgraph, while significantly decreasing the success rate of an adversary from near-perfect under \adgraph to around 8\% under \name.
Finally, we show that \name remains robust to a more sophisticated adversary that uses evasion techniques beyond those currently deployed on the web.

\end{abstract}

\section{Introduction}
\label{sec:introduction}

Users rely on privacy-enhancing blocking tools to protect themselves from online advertising and tracking.
Many of these tools---including uBlock Origin \cite{ubo_web}, Ghostery \cite{ghostery_web}, Firefox \cite{firefox_cookie_policy,firefox_redirect_tracking}, Edge \cite{edge_tracking_protection}, and Brave \cite{brave_tracking_protection}---rely on manually curated filter lists \cite{easylist,easyprivacy,disconnect_web} to block advertising and tracking.
The research community is actively developing machine learning (ML) approaches to automate the detection of advertising and tracking and make filter lists more comprehensive.
The first generation of ML-based blocking approaches analyze network requests \cite{Bhagavatula14MLforFilterListsAISec,Gugelmann15ComplementBlacklistPETS,Shuba2018PETSNoMoAds} or JavaScript code \cite{Wu16MLTrackingESORICS,Kaizer16JSTrackingIWSPA,Ikram17SeamlessTrackingPETS} to learn distinctive behaviors of advertising and tracking. 
However, these ML-based blocking approaches are highly susceptible to adversarial evasion techniques that are already found in the wild, including URL obfuscation \cite{Snyder20WhoFilterstheFiltersSigmetrics} and code obfuscation \cite{Skolka2019ObfuscationWWW}. 
To address this limitation, the next generation of ML-based blocking approaches leverage cross-layer graph information from multiple layers of the web stack~\cite{Iqbal20AdgraphSP,Sjosten20UnderservedRegionsWWW}.
These approaches claim better robustness to evasion, as compared to single-layer approaches, due to their use of \textit{structural} features of the graph (i.e., the hierarchy of resource inclusions) in addition to traditional \textit{content} features (i.e., the resource's network location or response content).

In this paper, we show that state-of-the-art ad and tracker detection approaches, such as \adgraph \cite{Iqbal20AdgraphSP}, are susceptible to adversarial evasion due to their disproportionate reliance on easy-to-manipulate content features. 
We show that a third-party adversary can achieve 8\% evasion success by manipulating URLs of its resources.
Worse yet, an adversary can achieve near-perfect evasion---as high as a 96\% success rate---if they collude with the first party, e.g, by using the CNAME cloaking technique already deployed by some trackers~\cite{Dao20CNAME,Dimova21CNAMEPETS}.

We introduce \name, the first ML-based ad and tracker blocking approach that \emph{does not rely on content features}.
\name improves the cross-layer graph representation by capturing a fundamental property of advertising and tracking services (ATS): the flow of information from one entity to the browser's storage, the network, and to other entities loaded on a page.
The intuition behind adding these features is to focus on the \textit{actions} of the advertising and tracking services, rather than the \textit{contents} of their resources.
We posit that actions are harder to obfuscate.
Advertising and tracking scripts need to generate and store identifiers for users, and those identifiers must be shared with any other entity with which they wish to share data (e.g., via cookie syncing \cite{Papadopoulos19cookiesyncing}).
Ultimately, if a script wishes to store an identifier in the browser, it will need to call a browser API, and as such, we monitor the flow of information to and from browser APIs. 
We build a graph representation of the page load by monitoring network requests, JavaScript execution, HTML element creations, and browser storage access.
From this graph we extract \textit{flow} features, which explicitly capture distinctive information flows in advertising and tracking. 
Our evaluation shows that \name's graph representation and flow features can entirely supplant content features, with comparable accuracy.

While high accuracy is necessary for deployment, it is not sufficient.
We have repeatedly seen that advertisers and trackers will attempt to circumvent detection and evade blocking \cite{Snyder20WhoFilterstheFiltersSigmetrics,Skolka2019ObfuscationWWW,Dao20CNAME}.
Therefore, in order for an advertising and tracking classifier to be useful in practice, it must be robust to adversarial manipulation.
We show that \name represents a significant step forward in robustness to adversarial evasion when compared to previous approaches.
In particular, we find that \name is robust to the types of URL, CNAME, and content manipulation evasion techniques that are in use on the web today.
We also know that ad and tracking adversaries will attempt to deploy more sophisticated evasion techniques tailored to our classifier.
To understand how robust \name would be in the face of these new evasion techniques, we propose a novel realistic graph manipulation evasion technique.
We show that this attack achieves only limited evasion success against \name, while incurring a non-trivial usability loss in terms of mistakenly blocking its own advertising/tracking resources or other benign resources on the web page.

Overall, our findings suggest that the community should migrate away from unreliable content features for advertising and tracking blocking.
We show that information flow features built upon the actions of advertisers and trackers provide a promising path forward.

In summary, our contributions are as follows:

\vspace{-2mm}

\begin{itemize}
  \setlength\itemsep{-.3em}

\item We show that existing ML-based ad and tracker detection approaches are susceptible to evasion due to their heavy reliance on content features. As a representative example, we show how an adversary can achieve near-perfect evasion of \adgraph using evasion techniques already in use on the web today.

\item We introduce \name, the first ML-based ad and tracker blocking approach that does not rely on content features and captures fundamentally distinctive information flows in advertising and tracking.  

\item Our in-depth evaluation shows that \name achieves comparable accuracy to prior approaches and achieves significantly better robustness to adversarial manipulation of content features.

\item We propose a novel graph manipulation evasion technique, and show that \name (and the information flow features it relies on) remain robust under this sophisticated attack. 

\end{itemize}

\para{Paper organization:} 
The rest of this paper is organized as follows: 
Section~\ref{sec:background} provides an overview of recent advance in ML-based ad and tracker blocking. 
Section~\ref{sec:adgraph} evaluates robustness of existing graph-based approaches, using \adgraph as a representative example.
 Section~\ref{sec:dataflow} describes the design and evaluation of \name. 
 Section~\ref{sec:robustness} further evaluates \name's robustness to adversarial attacks. 
We discuss limitations of our work in Section~\ref{sec:improvements} and conclude in Section~\ref{sec:conclusion}.

\section{Background \& Related Work}
\label{sec:background}

Online behavioral advertising enables ad targeting based on users' interests and behaviors. 
To target ads, online advertising relies on the intertwined tracking ecosystem that uses cookies for cross-site tracking. 
For instance, the real-time bidding (RTB) protocol that powers programmatic online advertising has built-in mechanisms for advertisers and trackers to share information \cite{Fouad20PixelsPETS,Papadopoulos19cookiesyncing}. 
Thus, almost always, ads and trackers go together, often with intertwined execution flows and resource dependencies.
Below, we revisit prior literature on ad and tracker blocking, and analyze its limitations.

Popular ad and tracker blocking tools such as Adblock Plus \cite{adblockplus_web} rely on filter lists \cite{easylist,easyprivacy}.
These filter lists are manually curated based on user feedback. 
Prior work has shown that manually curated filter lists suffer from \textit{scalability} and \textit{robustness} issues. 
First, filter lists have trouble keeping up with the ever expanding advertising and tracking ecosystem. 
Filter lists have grown to include tens of thousands of rules that are often not updated in a timely fashion. 
For instance, prior work showed that filter lists may take as long as 3 months to add rules for newly discovered ads and trackers \cite{Iqbal17AntiABIMC}.
Once a filter rule is added to block an advertising and tracking service, it is rarely removed, even if it is no longer needed. 
In fact, prior work showed that almost 90\% of the rules in filter lists are rarely or never used \cite{Snyder20WhoFilterstheFiltersSigmetrics}.
Second, filter lists are not robust to evasion attempts by advertisers and trackers. 
Filter lists are brittle in the face of domain rotation \cite{dga_blog,Zaifeng2018DGAunblock} and manipulation of page structure \cite{facebookadblocking,facebook_newmanipulation,facebook_html_rand}.
For instance, prior work showed that filter lists are susceptible to evasion attacks such as randomization of URL path, hostname, or element attributes and IDs \cite{Wang16FSCWebRanz,Alrizah19errorsfilterlists}.

\para{Addressing scalability.} 
To address the scalability issues that arise due to manual curation of filter lists, researchers have proposed to use machine learning (ML) for automated ad and tracker blocking.
Prior ML-based approaches mainly detect ads and trackers at the network and JavaScript layers of the web stack. 
Specifically, these approaches detect ads and trackers by featurizing network requests \cite{Bhagavatula14MLforFilterListsAISec,Gugelmann15ComplementBlacklistPETS,Shuba2018PETSNoMoAds} or JavaScript code \cite{Wu16MLTrackingESORICS,Kaizer16JSTrackingIWSPA,Ikram17SeamlessTrackingPETS}.

Network layer approaches rely on content in URLs, HTTP headers, and request and response payloads (e.g., keywords, query strings, payload size) to extract features and train ML models to detect ads and trackers \cite{Bhagavatula14MLforFilterListsAISec,Gugelmann15ComplementBlacklistPETS}. 
While trying to mimic filter lists by detecting ad and tracker URLs, these approaches end up replicating some characteristics of filter lists and thus also naturally inherit their shortcomings.
For example, presence of a certain keyword in the request URL could be a distinguishing feature.
However, as discussed earlier, such keyword based features are brittle in the face of trivial evasions such as domain rotation \cite{Wang16FSCWebRanz,Alrizah19errorsfilterlists}.

JavaScript layer approaches rely on static or dynamic analysis to extract features and train ML models to detect ads and trackers.
Examples of features are n-grams of code statements obtained via static analysis \cite{Ikram17SeamlessTrackingPETS} or JavaScript API invocations captured via dynamic analysis \cite{Wu16MLTrackingESORICS}. 
These approaches are susceptible to JavaScript obfuscation \cite{Dang2017EvadingPDFJSClassifiers,Fass2019HideNoSeek,Hansen2020AttackJavaScriptClassifiers}.
These approaches are also susceptible to evasion such as script amalgamation or dispersion. 
They implicitly assume that tracking code is bundled in a single script or that tracking scripts only contain tracking code.
However, in practice,  tracking code could be distributed across several chunks and packaged with functional code \cite{Iqbal20AdgraphSP}.

\para{Addressing robustness.}
While network and JavaScript layer approaches consider information at each layer in isolation, ads and trackers rely on all three layers (i.e. network, JavaScript, and HTML) of the web stack for their execution.
Therefore, it is natural that focusing on only one layer lacks robustness against the aforementioned evasion attempts.
To address this limitation, graph-based approaches aim to capture the interactions among and across network, JavaScript, and HTML layers of the web stack.

Graph-based approaches extract features from the cross-layer graph representation to train ML models to detect ads and trackers \cite{Iqbal20AdgraphSP, Sjosten20UnderservedRegionsWWW}. 
These approaches leverage rich cross-layer context and thus claim to be robust to evasion attempts. 
\adgraph was the first graph-based approach to ad and tracker classification \cite{Iqbal20AdgraphSP}. 
It extracts structural features from the graph such as node connectivity and ancestry information as well as content features such as URL length and presence/absence of certain keywords.
Sjösten et al.~\cite{Sjosten20UnderservedRegionsWWW} introduced PageGraph, which extends \adgraph's graph representation by improving event attribution and capturing more behaviors. 
In addition to content and structural features, they also added perceptual features to train the classifier. 
Since \textit{perceptual} features attempt to use the rendered resource content, they are also considered content features. 
Chen et al.~\cite{Chen2021SignaturesSP} proposed an approach, using PageGraph, to detect trackers based on their execution signatures.
In contrast to ML-based approaches, their signature-based approach would only be able to detect trackers that strictly match the signatures of tracking scripts, but miss trackers with even slight deviations in their behavior, such as changes in the execution order. 
Kargaran et al.~\cite{Kargaran20DependencyGraph} followed a different approach. 
Instead of building a graph representation per website, they combined graph representations across multiple websites to model relations between third parties on those sites.
Just like \adgraph, they also extract structural and content features from the graph to train the classifier.
These graph-based systems use a combination of content and structural features for classification, which they claim increases the robustness to evasion attacks. 
While this combination should intuitively improve classifier robustness, we posit that it would be less robust than expected if the classifier relies heavily on content features. 
This is because content features pertain to a single node on the graph and are easy to manipulate for an adversary, e.g., using adversarial attacks on textual \cite{Zhu20A4} and perceptual \cite{Tramer19CCSAdVersarial} content features, without causing undesired changes in other nodes. 
It is noteworthy that Zhu et al.~\cite{Zhu20A4}, also manipulate structural features, however their manipulations are only limited to graph size. 
Further, they do not evaluate the impact of their mutations on overall graph.

In the next section, we analyze the robustness of graph-based ad and tracker detection systems. 
We focus on \adgraph as it is representative of other graph-based systems that use similar structural and content features.

\section{\adgraph Robustness}
\label{sec:adgraph}

\newcommand{\RomanNumeralCaps}[1]
    {\MakeUppercase{\romannumeral #1}}

In this section, we analyze \adgraph's robustness by evaluating its accuracy in the face of adversarial content manipulation.

\adgraph is a graph-based machine learning approach that detects ads and trackers based on their structural and content properties. 
\adgraph instruments the Chromium web browser to capture detailed execution of ads and trackers across the HTML, JavaScript, and the network layer, and models the interaction among these layers in the form of a graph. 
Using this graph, \adgraph extracts two categories of features: \textit{content} (information related to individual nodes in the graph, such as URL length and presence of ad/tracking keywords in the URL) and \textit{structure} (information about relationships between nodes, such as connectivity and ancestry information).
It uses the extracted features to train a machine learning classifier to detect advertising and tracking resources. 
The full list of \adgraph features are described in Table \ref{tab:dataflow_features}.

Since \adgraph relies on content properties, in addition to structural properties, it is subject to same evasion attacks that succeed against the filter lists-based ad and tracker detection approaches \cite{Wang16FSCWebRanz,Alrizah19errorsfilterlists}. 

\subsection{Threat Model \& Attack}
Our threat model assumes an adversarial third-party advertiser or tracker embedded on a site, who aims to change the classification of its resources from advertising and tracking services ({ATS}) to benign resources ({Non-ATS}) in order to evade detection by ad and tracker blocking tools. 

We assume that the adversarial third party has limited cooperation with the first-party publisher.
We do not assume full cooperation because the parties are mutually distrusting.
The third-party adversary generally does not trust the first-party publisher to serve its advertising and tracking resources via a reverse proxy \cite{bypassing_adblockers_ga,tracking_visitors_with_adblockers}.
Likewise, the first-party publisher does not trust the third-party adversary to host functional resources via the adversary-controlled CDN \cite{adblocking_drives_disruptions}. 
Given existing practices, we assume that the adversary can serve its advertising and tracking resources from a first-party subdomain but not arbitrarily within the first-party domain space.
For example, the adversary can masquerade its resources through CNAME cloaking ~\cite{cnamecloaking2019}, which only requires a minor change in DNS records by the first party.
Recent measurement studies have reported an increase in the prevalence of CNAME cloaking over the last few years. 
Dao et al. \cite{Dao20CNAME} showed that the usage of CNAME cloaking-based tracking has steadily increased between 2016 and 2020, with 1,762 of Alexa's top-300K websites employing at least one CNAME-based tracker as of January 2020.
Dimova et al. \cite{Dimova21CNAMEPETS} also showed that the usage of CNAME cloaking has increased by 22\% from 2018 to 2020, with 9.98\% of Tranco's top-10K websites now employing at least one CNAME-based tracker as of October 2020.

We assume that the adversary is able to manipulate their own URLs by altering the domain name or query string. 
Naturally, the adversary can only manipulate URLs that are under their control, and only attempts to manipulate the subset of its URLs that were initially correctly classified as {ATS} (ad and tracker URLs initially classified as {Non-ATS} already benefit the adversary).
The adversary cannot manipulate the data used to train the classifier.
Therefore, we only implement mutations during inference.
We implement two types of URL manipulations. 
For domain names, we allow the adversary to randomly change the URL's domain, subdomain, or both.
In practice, adversaries can rely on automated techniques to generate random domains and subdomains.
For example, they can use malware-inspired domain generation algorithms (DGA) techniques to generate a large number of domains \cite{dgaBlog,Plohmann16DGA}.
%
For query strings, we randomly change the number of parameters, the parameter names, the parameter values in the URL, or a combination of the three.

\begin{figure}[!t]
\centering
\includegraphics[width=1\columnwidth,trim=0.0cm 2.0cm 0.0cm 0.0cm]{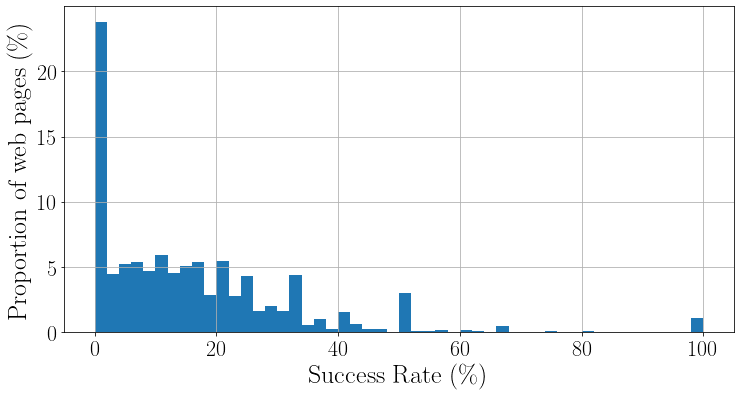}
\caption{\looseness=-1 Classification switch success rate distribution by web page (over 10 folds) when the adversary does \textbf{not} collude with the first party. The average success rate per web page is 15.92 $\pm$ 0.03 \%.} 
\label{fig:success_rates}
\vspace{-5pt}
\end{figure}

\subsection{Results}
\label{sec:evaluation_adgraph}

\para{Experimental setup.} We extend OpenWPM \cite{Englehardt16OpenWPMCCS} to automatically crawl websites with Firefox and build \adgraph's representation.
We crawl 10K sites sampled from the Alexa's top-100K list, the top 1K sites and a random sample of 9K sites ranked between 1K-100K, and store their graph representations.
Next, we implement a decision tree classifier that closely follows \adgraph's design \cite{Iqbal20AdgraphSP}, and extract features from the graphs for training and testing.
For ground truth, we use the same set of filter lists for data labeling that were used by \adgraph \cite{Iqbal20AdgraphSP}. 
A URL is labeled as {ATS} if it is present in one or more of the filter lists, and {Non-ATS} otherwise. 
We use 10-fold cross validation to obtain our results, where the folds are selected such that every fold uses a different set of web pages in the test set. 
Our classifier obtains comparable performance to the original results reported by \cite{Iqbal20AdgraphSP}: 92.33\%  accuracy, 88.91\%  precision, and 92.14\% recall.
The minor differences are likely due to differences in crawled sites, updated filter lists, and a few subtle changes in our adaptation of \adgraph from online to offline.
In \adgraph's online implementation, features are extracted from each node in the graph as they are created. 
Our offline adaptation, instead, extracts features after page load completion. 
There are also some minor differences due to JavaScript attribution, caused by the differences in instrumentation between Chromium-based \adgraph and Firefox-based OpenWPM.\footnote{Due to these differences, our features are not exactly identical to the online implementation of \adgraph. 
For example, in \adgraph, a node can have a maximum of two parents, which need not be the case for our system. 
Therefore, we do not use \adgraph features specific to these two parents. 
The full feature list, showing these differences is provided in Appendix~\ref{sec:adgraph-comparison}.}

\para{Adversarial success rate without collusion.}
In our first experiment, we assume that the adversary does not collude with the first party. 
The adversary can randomize their domain and subdomain, but cannot masquerade as the first party.
Our content mutation procedure results in the mutation of 41.48 $\pm$ 1.47 \% of all the test data URLs (averaged over 10 folds). 
The adversary's success rate in evading the classifier is 8.72 $\pm$ 0.42 \% (over 10 folds).
While this may seem like a low percentage, we note that every successful mutation is a win for the adversary since it means that one more of their ads or trackers is now unblocked.
Over all 10 folds, the adversary mutated 691,602 URLs, out of which 60,270 had their classifications switched. 

We also observe that the evasion success rate varies across sites, as shown in Figure~\ref{fig:success_rates}.
For $\approx$1\% of the web pages in the test set (90 pages), the adversary achieves a perfect success rate, meaning that all third-party ads and trackers on the web page are now classified as benign content. 
It is noteworthy that 21.62\% of the unblocked URLs belong to popular ad exchanges, which are responsible for further diffusion of user information due to the broadcast nature of real-time bidding (RTB) ~\cite{Bashir18DiffusionPETS}.
These unblocked ad exchanges can amplify the privacy harm because they often share information about page visits with multiple advertisers and trackers.

\begin{figure}[!t]
\centering
\includegraphics[width=1\columnwidth,trim=0.0cm 2.0cm 0.0cm 0.0cm]{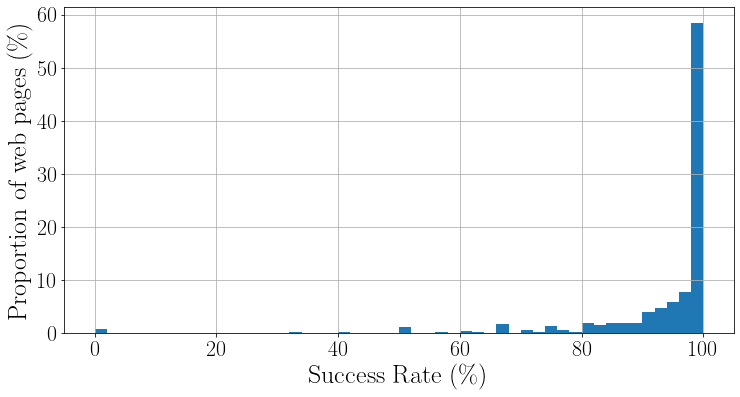}
\caption{\looseness=-1 Classification switch success rate distribution by web page (over 10 folds) when adversary colludes with the first party. The average success rate per web page is 93.01 $\pm$ 0.01 \%.} 
\label{fig:success_rates_first_party}
\vspace{-5pt}
\end{figure}

\para{Adversarial success rate with collusion.}
In our second experiment, we assume that the adversary colludes with the first party. 
The adversary can perform domain mutation such that their URL is a subdomain of the first party. 
The adversary's success rate increases to 96.62 $\pm$ 0.37 \% (over 10 folds). 
This means that being able to use a first-party subdomain provides almost perfect evasion capabilities. 
Figure~\ref{fig:success_rates_first_party} shows the evasion success rate variation across sites. 
For $\approx$50\% of the web pages in the test set, the adversary achieves a perfect success rate.
We also see a higher proportion (32.25\%) of the unblocked URLs belonging to popular ad exchanges, as compared to the previous experiment.
To better understand why such URL manipulation is able to evade detection by \adgraph, we analyze feature importance using information gain (see Table~\ref{tab:importance}).
We see that content features are essential to the \adgraph classifier: not only are the top-3 most important features content features, their relative importance scores are also high compared to the other features. 
Two of the top-3 features depend on whether a URL is third-party, which explains why we obtain high success rates when the adversary has the capability to masquerade as the first party.
These two features do not have an effect in the case where the adversary does not collude with the first party, since the adversary cannot change the fact that they are third party. 
However, the adversary's manipulations still influence the third top feature, length of the URL. 
Hence, we observe lower but non-trivial success rates even without collusion.

\begin{table}[!t]
  \centering
  \resizebox*{\columnwidth}{!}{%
  \begin{tabular}[c]{  l  l  l  }
   \toprule
     \textbf{Feature} & \textbf{Category} & \textbf{Information Gain (\%)} \\
    \midrule
    URL length & Content & 14.87 $\pm$ 0.36 \\   
    URL domain is a subdomain of the first party & Content & 11.06 $\pm$ 1.24 \\  
    URL is a third party & Content & 10.67 $\pm$ 1.32 \\
    Degree of a node &  Structure & 7.56 $\pm$ 0.63 \\
	Number of edges divided by number of nodes & Structure & 7.48 $\pm$ 0.41  \\
    \bottomrule
  \end{tabular}}
   \caption{Top 5 most important features for \adgraph's classification, their category, and information gain values (averaged over 10 folds). 
   }
  \label{tab:importance}
  \vspace{-15pt}
\end{table}

These results show that graph-based classifiers such as \adgraph are vulnerable because of their over-reliance on content features.
In the next section, we propose an approach to improve the robustness of graph-based ad and tracker blocking tools.

\section{\name}
\label{sec:dataflow}

\label{subsec:infoflow}
Online advertising and tracking fundamentally relies on information sharing.
Trackers need to share information with each other to improve their coverage of users' browsing history \cite{Englehardt16OpenWPMCCS,Papadopoulos19cookiesyncing}. 
Trackers also need to share information with each other as part of built-in dependencies in programmatic advertising protocols \cite{Olejnik14SellingPrivacyNDSS,Acar14CCSTheWebNeverForgets}.
We contend that leveraging such fundamental information sharing patterns can help build accurate and robust classifiers for ad and tracker blocking. 
We introduce \name, a classifier that explicitly captures these information sharing patterns as part of its cross-layer graph representation of the execution of a web page.

\begin{figure*}[!t]
    \centering
    \subfigure[]
    {
    		\begin{adjustbox}{varwidth=0.9\columnwidth,fbox}
        \includegraphics[width=0.64\columnwidth]{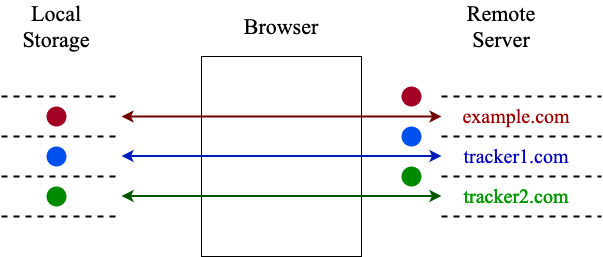}
                \label{fig:sharing_1}
                  \end{adjustbox}
    }
    \subfigure[]
    {
    		\begin{adjustbox}{varwidth=0.9\columnwidth,fbox}
        \includegraphics[width=0.64\columnwidth]{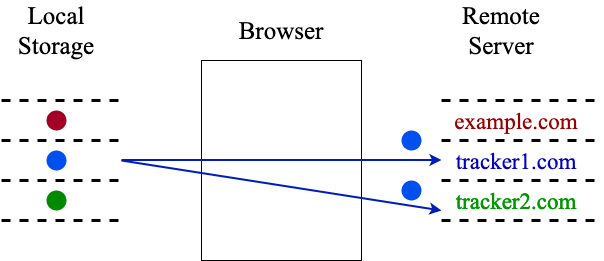}
                \label{fig:sharing_2}
                  \end{adjustbox}
    }
    \subfigure[]
    {
    		\begin{adjustbox}{varwidth=0.9\columnwidth,fbox}
        \includegraphics[width=0.64\columnwidth]{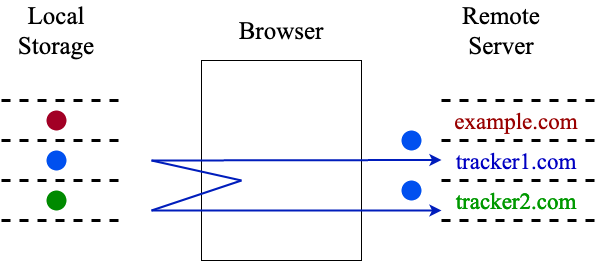}
        \label{fig:sharing_3}
        \end{adjustbox}
    }
    \caption{Origin isolation vs. sharing. Circles represent information about a user gathered by a particular domain (\texttt{example.com}, \protect\tikz{\protect\draw[fill=red, line width=0.1pt]  circle(0.6ex);}; \texttt{tracker1.com}, \protect\tikz{\protect\draw[fill=blue, line width=0.1pt]  circle(0.6ex);}; and \texttt{tracker2.com}, \protect\tikz{\protect\draw[fill=green, line width=0.1pt]  circle(0.6ex);}). The box  represents the browser which acts as channel between the local storage on the user's device and the remote server of each domain.
    ~\ref{fig:sharing_1} Illustrates origin isolation in the browser: every domain can only access information in their own storage. ~\ref{fig:sharing_2} and ~\ref{fig:sharing_3} illustrate two information sharing patterns that trackers use to circumvent origin isolation: (b) cookie syncing, where users' identifiers are sent to more than one domain; and (c) sharing identifiers using JavaScript APIs.}
    \label{fig:sharing}
    \vspace{-10pt}
\end{figure*}

To illustrate the information sharing patterns that we want to capture in \name, let us revisit how information sharing between different origins is mediated by the browser. 
We deliberately use a loose definition of origin.
An origin can be, depending on the specific use case, a site, a domain, or an entity, among others.
At a high-level, the web browser isolates different origins, based on various policies, so that their data is not leaked to each other.
Figure \ref{fig:sharing_1} illustrates how the browser limits information sharing between different origins: 
\texttt{example.com}, \texttt{tracker1.com}, and \texttt{tracker2.com} each have access to their isolated local storage (e.g., cookies, IndexedDB) that may be used to store user identifiers.
The browser isolates information flows between the local storage and remote servers of different origins:
\texttt{tracker1.com} and \texttt{tracker2.com} cannot generally access each others' cookies. 

Trackers typically circumvent these limitations in the browser in two main ways. 
First, Figure \ref{fig:sharing_2} illustrates how a tracker may share its identifier with another tracker through cookie syncing. 
This can be implemented in several ways.
For example, let's say \texttt{example.com} loads a JavaScript from Tracker~1 that first uses \texttt{document.cookie} to retrieve Tracker~1's identifier cookie from its cookie storage and then initiates a GET request to Tracker~2. 
The script includes Tracker~1's identifier cookie in the request URL as a query string parameter. 
Note that the request automatically includes Tracker~2's identifier cookie in the Cookie header.
Therefore, when Tracker~2's remote server receives the request, it would be able to sync Tracker~1's identifier with its own identifier. 
As another example, let's say \texttt{example.com} first loads an invisible pixel from Tracker~1, which responds back with a 3XX redirect status code along with the URL in the Location header that points to Tracker~2 and includes Tracker~1's identifier cookie.
Upon receiving the response, the browser issues a GET request to Tracker~2 and includes Tracker~1's identifier cookie in the request URL and Tracker~2's identifier cookie in the Cookie header. 
Again, Tracker~2's remote server is able to sync Tracker~1's identifier with its own identifier.

Second, Figure \ref{fig:sharing_3} illustrates how a tracker may share its identifier with another tracker through various JavaScript APIs, in several ways.
For example, let's say \texttt{example.com} loads scripts from Tracker~1 and Tracker~2 which then share their identifiers by reading/writing to the global variables of the \texttt{window} object.
The script from Tracker~1 may assign its identifier to a new global variable \texttt{foo} that is then read by the script from Tracker~2.
Therefore, Tracker~1 and Tracker~2's scripts would be able to sync identifiers with each other and also send them to their respective remote servers. 
As another example, let's say \texttt{example.com} loads iframes from Tracker~1 and Tracker~2 which then share their identifiers using \texttt{postMessage}.
While these iframes have different origins, Tracker~1's iframe can use \texttt{window.parent} property to get a reference to the parent window and then use \texttt{window.frames} to get a reference to Tracker~2's iframe.
Tracker~1's iframe can then use this reference to call \texttt{window.postMessage} and send its identifier to Tracker~2's iframe, which can use \texttt{window.addEventListener} to receive the identifier. 
Tracker~2's iframe can then send the shared identifier with its remote server to sync them. 
Trackers use a wide variety of information sharing patterns, beyond the two aforementioned mechanisms. 
A sound and precise examination of all patterns warrants full-blown information flow tracking that adds significant  implementation overheads and complexity \cite{Hedin14JSFlowSAC, Chudnov2015FlowCCS, Chen18MystiqueCCS}.
As we discuss next, \name approximately\footnote{See Section \ref{sec:improvements} for a discussion of completeness of \name's implementation.} captures these information sharing patterns by including additional nodes and edges in its graph representation that correspond to elements and actions associated with these information sharing patterns.
It then extracts new features on this enriched graph representation to train a classifier for detecting ads and trackers.

\subsection{Design \& Implementation}

\subsubsection{Graph Construction} 
\label{subsec:graph_construction}
\name captures the flow of information among and across the HTML, network, JavaScript, and storage layers of the web stack.
At the HTML layer, \name captures creation and modification of all HTML elements, e.g., \texttt{iframe}, that are initiated with scripts. 
At the JavaScript layer, \name captures the scripts' interaction with other layer, e.g., initiation of a network request.
At the network layer, \name captures all outgoing network requests and their responses. 
At the storage layer, \name captures read/write in cookies and local storage through scripts and network requests, and also the exchange of values between network requests. 

\textbf{OpenWPM Instrumentation.}
We extend OpenWPM \cite{Englehardt16OpenWPMCCS} to capture the execution and interaction of HTML, network, JavaScript, and storage layers. 
To capture HTML elements creation and modifications, we instrument \texttt{createElement} method and register a \texttt {MutationObserver} interface. 
To capture network requests, we parse OpenWPM's existing instrumentation, which uses a webRequests listener \footnote{https://developer.mozilla.org/en-US/docs/Mozilla/Add-ons/WebExtensions/API/webRequest}, to capture all of the network requests, their responses, and redirects. 
To capture JavaScript interaction, we parse OpenWPM's existing instrumentation, which relies on JavaScript's stack trace to log JavaScript execution. 
To capture read/write to storage, we instrument \texttt{document.cookie} and localStorage methods and also intercept cookie read/write HTTP headers. 
\textbf{Graph Composition.}
Elements at each of the layers are represented with nodes and the interaction between these nodes is represented with edges. 
Specifically, each HTML element, network request, script, and stored value, is represented as a node.  
Edges to HTML nodes from script nodes represent the creation and modification of elements. 
Edges from HTML nodes to network nodes represent initiation of network requests to load content, such as scripts and images. 
Edges from script nodes to network nodes represent the initiation of \texttt{XMLHTTPRequest} which will be parsed by the script. 
Edges between script and storage nodes and network and storage nodes, represent the read/write of values in the storage. 
Edges between network nodes either represent redirects or the presence of the same stored values.\footnote{We match stored values with their encoded and hashed counterparts. Specifically, we look for presence of base64 encoded and MD5 and SHA-1 hashed values \cite{Englehardt18EmailPETS, Fouad20PixelsPETS}.} 
\textit{Graph Composition Example.}
To illustrate \name's graph representation, let us consider the example web page given by Code \ref{code_snippet}.
The web page embeds a script from Tracker~1 and an iframe from Tracker~2.
The tracking iframe from Tracker~2 reads its tracking cookies and sends them to Tracker~3 via an XHR.
Both trackers trigger requests to share tracking identifiers.
The HTTP requests and responses that result from loads in Code \ref{code_snippet} are listed in Listing \ref{req_resp}.

\renewcommand{\lstlistingname}{Code}
\begin{lstlisting}[style=htmlcssjs, captionpos=b, label={code_snippet}, caption={An example web page sending requests to several trackers.},captionpos=b, numbers=left, xleftmargin=2em, float]
<html>
     <script src='tracker1.com/track.js'>
      ...
      var image =document.createElement('img');
      image.src = 'tracker2.com/sync';
      document.body.appendChild(image);
      ...
     </script>
     ...
     <iframe src='tracker2.com/track.html'>
      <script>
      ...         
      idCookie = document.cookie;
      var newReq = new XMLHTTPRequest();
      newReq.open("GET", "tracker3.com?user_id=" + idCookie);
      ...
    </script>
  </iframe>
</html>
\end{lstlisting}

\lstdefinestyle{myCustomStyle}{
  tabsize=4,
  showspaces=false,
  showstringspaces=false,
  escapechar={|}
}

\lstset{basicstyle=\footnotesize,style=myCustomStyle}

\renewcommand{\lstlistingname}{Listing} 
\setcounter{lstlisting}{0}
\begin{lstlisting}[style=htmlcssjs, captionpos=b, label={req_resp}, caption={HTTP requests and responses initiated from Code \ref{code_snippet}.}, captionpos=b, float]
---------------------------------------------------
|\color{blue}Request 1|
URL: tracker2.com/sync
Cookie: user1
|\color{blue}Response 1|
Status: 302
Location: tracker1.com?tracker2_id=user1
---------------------------------------------------
|\color{blue}Request 2|
URL: tracker1.com?tracker2_id=user1
|\color{blue}Response 2|
Status: 200
Set-Cookie: userA
Content: pixel.png
---------------------------------------------------
|\color{blue}Request 3|
URL: tracker3.com?user_id=user1
|\color{blue}Response 3|
Status: 200
\end{lstlisting}

\begin{figure*}[!htpb]
  \centering
  \subfigure[Graph representation of Code \ref{code_snippet} in \adgraph]
  {
      \includegraphics[width=0.8\columnwidth]{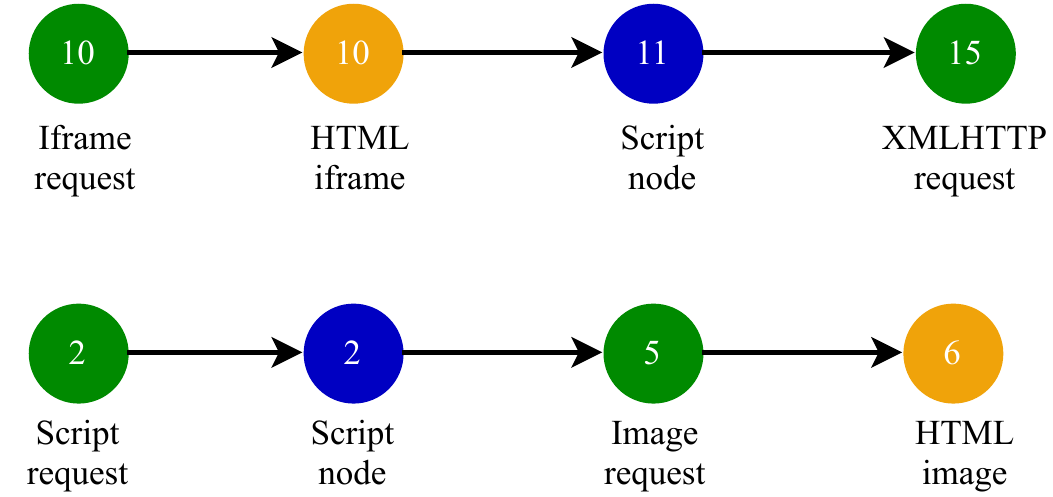}
              \label{fig:adgraph_representation}
  }
  \hfill
  \subfigure[Graph representation of Code \ref{code_snippet} in \name]
  {
      \includegraphics[width=\columnwidth]{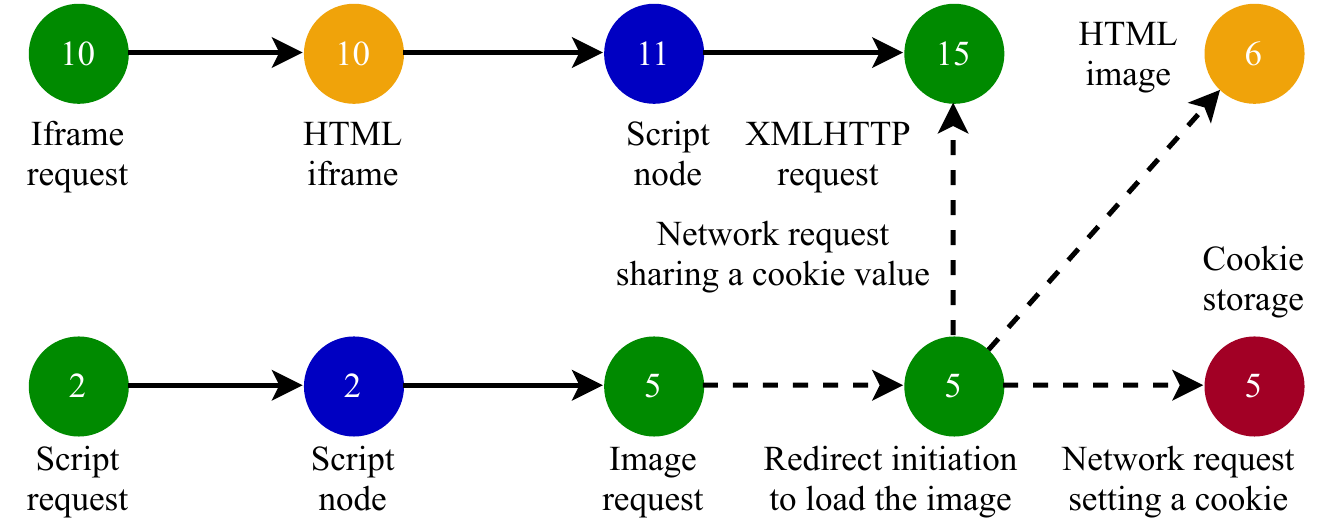}
      \label{fig:webgraph_representation}
  }
  \caption{Graph representation of Code~\ref{code_snippet} in \adgraph and \name. \protect\tikz{\protect\draw[fill=black!40!green] circle(1.0ex);} represents network nodes, \protect\tikz{\protect\draw[fill=black!40!blue] circle(1.0ex);} represents script nodes, \protect\tikz{\protect\draw[fill=black!10!yellow] circle(1.0ex);} represents HTML nodes, and \protect\tikz{\protect\draw[fill=black!40!red] circle(1.0ex);} represents storage nodes. Node numbers correspond to the lines in Code \ref{code_snippet}. In Figure \ref{fig:webgraph_representation}, dotted (- - -) lines represent the additional edges that are captured by \name and missed by \adgraph.}
  \label{fig:adgraph_webgraph_representation}
  \vspace{-10pt}
\end{figure*}

Tracker~1's script embeds an image element from Tracker~2, which causes the browser to send an HTTP request (Request~1 in Listing~\ref{req_resp}) that includes Tracker~2's cookie.
Tracker~2 responds to this request with a redirect to Tracker~1 that embeds the user identifier Tracker~2 received via the initial request's \texttt{Cookie} header (i.e., \texttt{user1}).
The browser makes a subsequent request (Request~2 in Listing \ref{req_resp}) to Tracker~1.
Tracker~1 responds with a tracking pixel image and a \texttt{Set-Cookie} header to set its own tracking cookie with the value \texttt{userA}.
On the backend, Tracker~1 knows that \texttt{userA} is known as \texttt{user1} by Tracker~2.
Tracker~2's embedded iframe further shares its identifier cookie with Tracker~3.
It does so by accessing its cookies locally via \texttt{document.cookie} and embedding them in an XHR to Tracker~3 (Request~3 in Listing \ref{req_resp}).

\textbf{Differences as compared to \adgraph.}
\name keeps \adgraph's HTML and JavaScript layers as they are, but extends the network layer and includes a new storage layer in the graph representation. 
\name also introduces information flow edges, which are absent in \adgraph, to entwine the extended network layer and the storage layer. 
The extension of network and the addition of storage layer allow \name to explicitly capture information sharing patterns used in advertising and tracking.

We illustrate the differences in Figure~\ref{fig:adgraph_webgraph_representation} which shows the graph representation of the web page in Code \ref{code_snippet} and request and response sequences in Listing \ref{req_resp} for both \adgraph (Figure~\ref{fig:adgraph_representation}) and \name (Figure~\ref{fig:webgraph_representation}). 
\adgraph's representation of the example web page consists in two disjoint graphs which capture the individual actions of the two trackers:
The first row of nodes (from 10 to 15) captures Tracker~2's tracking behavior: from the iframe loading to the initiation of an XHR request.
The second row of nodes (from 2 to 6) captures Tracker~1's tracking behavior: from the script loading to the initiation of a network request for loading an image.
In this figure, it becomes clear that \adgraph \emph{does not} capture the information sharing pattern between the nodes, because of its inability to capture the redirect (Request~2) made by the image request (network node 5) and the cookie set (storage node 5; visible only in \name's graph) by the redirect request.
\name, on the contrary, not only captures the flows appearing in \adgraph, but also captures the redirects (dotted edge between the two network nodes labeled 5) and cookies set by requests (the second network node 5 to storage node 5). 
This representation further enables \name to link requests that share common identifiers (node 5 to 15).

\subsubsection{Features}
\label{subsec:features}

We take the \adgraph feature set and augment them with three categories of features. These additional features come from  \name's improved graph representation, i.e., extension of the network layer and a new storage layer.
The features target storage, network, and information sharing behaviors that were absent in \adgraph. 
First, we extract features that measure the number of read/write cookie and localStorage accesses by a node. 
We obtain these features from the new storage layer. 
Second, we extract features that measure the number of requests and redirects to/from a node as well as the depth of a node in a redirect chain.
These features come from our extension to the network layer.
Third, we extract features that measure the number of different types of information sharing edges (e.g., nodes access the same storage node or share data of a storage node) to/from a node. 
We obtain these features using both the network and storage layers in \name's graph representation.
We also extract some standard graph features (e.g., in-degree, out-degree, eccentricity) for the information sharing edges. 
We jointly refer to these three newly added categories of features as \textit{flow} features.
Table \ref{tab:dataflow_features} in Appendix \ref{sec:adgraph-comparison} lists the full set of features in \name, including these newly added flow features.

\begin{figure*}[!htpb]
  \centering
  \subfigure[]
  {
      \includegraphics[width=0.64\columnwidth]{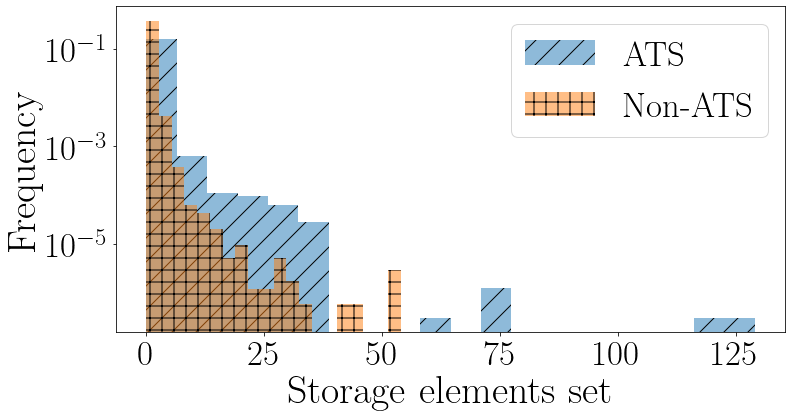}
              \label{fig:set_storage}
  }
  \subfigure[]
  {
      \includegraphics[width=0.64\columnwidth]{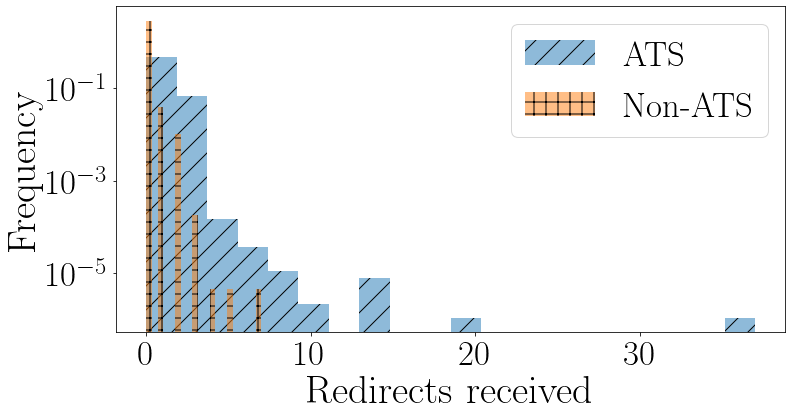}
              \label{fig:rec_redirects}
  }
  \subfigure[]
  {
      \includegraphics[width=0.64\columnwidth]{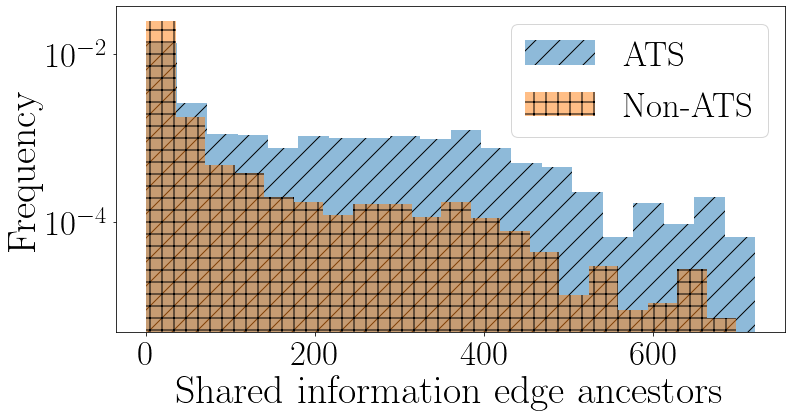}
      \label{fig:indirect_ancestors}
  }
  \caption{Histograms of three example flow features for ATS and Non-ATS resources (normalized, y-axis in log scale). (a) Number of storage elements set by a resource; (b) Number of network redirects received by a resource, and (c) Number of shared information ancestors of a resource. These features demonstrate different distributions for ATS and Non-ATS resources, and thus can help the classifier to distinguish between them.}
  \label{fig:feature_distribution}
  \vspace{-10pt}
\end{figure*}

To illustrate the potential of these features in distinguishing ATS and Non-ATS resources, let us consider three flow features belonging to each of the categories described above:
the number of storage elements set by a resource (Figure~\ref{fig:set_storage}),
the number of requests that were redirected to a resource (Figure~\ref{fig:rec_redirects}),
and the number of information sharing edge ancestors (Figure~\ref{fig:indirect_ancestors}).
As explained in Section~\ref{subsec:infoflow}, ATS resources store user identifiers in storage elements and use redirects and sharing of identifiers in URLs to perform actions such as cookie syncing.
Therefore, we expect ATS resources to set a larger number of storage elements, be at the receiving end of redirects, and be involved in a larger number of shared information edges than Non-ATS resources.
We plot in Figure~\ref{fig:feature_distribution} the distributions of these features in our dataset.
We see that, indeed, the distributions are different for benign and ATS resources, with ATS presenting higher values on average for the three features under study.
The differences in distributions is especially apparent for Figure~\ref{fig:indirect_ancestors}, which shows the number of shared information edge ancestors.
In our dataset, we observe 589,218 cases of ATS receiving a cookie value in a request URL, as compared to 89,564 cases for non-ATS.
This sharing is detected as an information sharing edge, which in turn  leads to ATS having larger values in shared information edge properties than Non-ATS.
In the case of redirects (Figure~\ref{fig:rec_redirects}), the probability that a Non-ATS resource has more than 7 redirects tends to 0, which is not the case with ATS resources.
The number of ATS resources with more than 7 redirects is very small in our dataset ($\approx$ 0.04\%).
Yet, it is a top-20 feature in our classifier, as observing more than 7 redirects directly identifies the resource as ATS.
Storage element setting (Figure~\ref{fig:set_storage}) shows a similar behavior, with ATS resources sometimes having more than 54 elements set, while Non-ATS resources never have so many.
While individual contributions of some of these flow features might be small, they provide a strong signal in distinguishing ATS when combined, as we show in the next section.

\subsection{Evaluation}
\label{sec:dataflow-results}
To evaluate \name, we use the same dataset of 10K web pages and method as in Section \ref{sec:evaluation_adgraph}. 
To understand the marginal benefit of \name over \adgraph, we systematically compare the performance of different feature sets and graph representations.
Table~\ref{tab:performance_webgraph} summarizes the results.

\begin{table}[htbp]
  \centering
  \resizebox*{\columnwidth}{!}{%
  \begin{tabular}[c]{ l l c c c }
   \toprule
     \textbf{Graph} & \textbf{Feature Set} & \textbf{Accuracy} & \textbf{Precision} & \textbf{Recall} \\
    \midrule
    \adgraph & Structural + Content & 92.33 $\pm$ 0.50 & 88.91 $\pm$ 1.14 & 92.14 $\pm$ 0.65 \\
     & Structural & 80.22 $\pm$ 0.81 & 71.85 $\pm$ 1.53 & 82.44 $\pm$ 1.26 \\
    \name & Structural+ Flow + Content & 94.32 $\pm$ 0.27 & 92.24 $\pm$ 0.67 & 94.14 $\pm$ 0.30 \\
         & Structural + Flow & 86.93 $\pm$ 0.64  & 80.57 $\pm$ 1.12 & 90.01 $\pm$ 0.50 \\
         & Structural  & 82.62 $\pm$ 0.47 & 75.67 $\pm$ 0.75 & 85.09 $\pm$ 1.41 \\
    \bottomrule
  \end{tabular}}
     \caption{Evaluation of \name and \adgraph with different feature set variations.} 
  \label{tab:performance_webgraph}
\end{table}

We observe that \adgraph's performance drops by at least 10\% when content features are removed.
Recall from Section \ref{sec:evaluation_adgraph} that if content features are present alongside structural features, \adgraph is particularly susceptible to evasion: trackers have an 8.72\% evasion success rate on their own, and a 96.62\% success rate if they collude with the first party.
Thus, there is a trade-off in \adgraph between effectiveness (with content) and robustness to evasion (without content).

Second, Table~\ref{tab:performance_webgraph} shows that \name's performance is better than \adgraph due to its improved graph representation and new flow features. 
When using all feature sets, \name outperforms \adgraph by about 2-4\%.
If we remove content features for robustness, we observe a drop in accuracy limited to just 4-9\% across all measures.
We conclude that \name's improved graph representation and new flow features can compensate for the loss of content features to a large extent.

Finally, Table~\ref{tab:performance_webgraph} shows that \name's improved graph representation by itself (i.e., even without the new flow features) contributes to about half of the improvement over \adgraph. 
\name with only structural features achieves 2-4\% improvement across all measures as compared to \adgraph also with only structural features. 
We conclude that, while \name's new flow features help improve its accuracy, the improved graph representation is an important contributor to performance.

\begin{table}[htbp]
  \centering
  \resizebox*{0.9\columnwidth}{!}{%
  \begin{tabular}[c]{  l l l }
   \toprule
     \textbf{Feature} & \textbf{Category} & \textbf{Information gain (\%)} \\
    \midrule
   Shared information ancestors & Flow  & 6.48 $\pm$ 0.69 \\
  Number of requests sent  by node &  Flow & 5.9 $\pm$ 0.69 \\
   Number of nodes in graph & Structure & 5.46 $\pm$ 0.35 \\
  Average degree connectivity of node & Structure & 5.18 $\pm$ 0.16 \\
  Number of edges in graph & Structure & 4.19 $\pm$ 0.34 \\
    \bottomrule
  \end{tabular}}
   \caption{Top-5 most important features for \name's classification, their category, and information gain (averaged over 10 folds).}
  \label{tab:importance_webgraph_nodomain}
  \vspace{-10pt}
\end{table}

To provide insights into the relative importance of flow and structural features, we list top five most important features in terms of information gain in Table \ref{tab:importance_webgraph_nodomain}.
The two most important features are flow features. 
As discussed in Section~\ref{subsec:features}, the top feature distribution (Figure~\ref{fig:indirect_ancestors}) is very different for ATS and non-ATS, so it's not surprising that this feature contributes to the classification.
Storage setting (Figure~\ref{fig:set_storage}) and received redirects (Figure~\ref{fig:rec_redirects}) contribute a smaller, but still useful, portion towards identification; they have information gains of 1.9\% ($\pm$ 0.37) and 2.5\% ($\pm$ 0.47) respectively (21st and 17th most important features).
We also observe that structure features, enhanced by \name's improved graph representation, also contribute towards the performance. 
We further analyze which features contribute most to each prediction of \name using \texttt{treeinterpreter}~\cite{treeinterpreter}.
For $\approx$ 32\% of predicted ATS in the dataset, the flow features were the top contributors, indicating that they provide an important signal for the presence of trackers.
In contrast, for $\approx$ 47\% of predicted Non-ATS in the dataset, structure features were the top contributors.
These results confirm our earlier intuition that capturing information sharing behaviors that are unique to advertising and tracking carries significant predictive power.
\subsection{Efficiency} 
We envision \name to be used for filter list curation and maintenance in an offline setting. 
\name relies on large scale web crawls and notoriously expensive graph traversals for feature extraction.
We now measure \name's offline overhead to demonstrate its adequacy as a tool to periodically update filter lists. 
\parait{Crawl time.} Our implementation of \name has an upper bound of 60 seconds, enforced with a timeout, to crawl a website.
In the average case, crawls take only $\sim$26.46 seconds. 
Crawls can be parallelized over several instances to reduce the crawl time. 
For example, it took us around $10.5$ hours to crawl 10K websites, parallelized over 7 instances. 
Without parallelization and if all websites would reach the timeout, the crawls would take $\sim$166 hours. 

\parait{Processing websites.} On average, \name takes 0.72 seconds to build the graph, 15 seconds to extract features, and 0.25 seconds to train and test each website.
For our crawl of 10K websites, it took us a total of $\sim$44.36 hours to create their graphs and extract features on a single instance.  
This time can be significantly reduced using parallelization.

\parait{Update frequency.}
These estimates suggest that for 10K websites containing $\sim$1.1 million requests \name will require, at most, $\sim$166 (data crawling) and $\sim$44.36 (data processing) hours with a single instance. 
However, when averaged over 7 instances, the computation time significantly reduces to only 16.83 hours (10.5 for crawling and 6.33 for processing).
We anticipate the computation time for periodic updates to reduce significantly because many websites have low update frequency. 
Specifically, monitoring the update frequency of websites will allow us to only crawl when changes are expected in websites. 
In cases, where we determine that the website did not change since the last crawl, we will not recompute their classifications.
With this performance, \name could be able to update filter lists on a daily basis, and certainly operate within the current the expiry period (mandated update frequency) of popular filter lists, e.g., 4 days for Easylist \cite{easylist}. 
Frequent updates with \name can help remove outdated rules and as well as add new rules to block newly discovered ads and trackers. 

\section{\name Robustness}
\label{sec:robustness}

In this section, we evaluate \name's robustness against content mutation attacks (described in Section~\ref{sec:adgraph}) and structure mutation attacks. 

\subsection{Content mutation attacks} 
To evaluate \name against content mutations, we strengthen the threat model described in Section~\ref{sec:adgraph} to enable the adversary to also masquerade their resources as first party, i.e., through first-party subdomains.
Overall, our attacks involve random mutations to domain names, subdomains, and the query string in URLs (Section~\ref{sec:evaluation_adgraph}). 

By relying on content mutations, the adversary is able to switch 96.62\% of their \texttt{ATS} resources to \texttt{Non-ATS} against \adgraph.
Against \name, the adversary's success rate plummets to just 8.34 $\pm$ 0.66\% (over 10 folds). 
For example, \texttt{mylivesignature.com}, a tracking domain, was able to switch all of its 560 \texttt{ATS} resources to \texttt{Non-ATS} against \adgraph, but none against \name.

Note that, even though \name does not use content features, the evasion success rate against \name does not drop to zero.
This is because some of the \name's features implicitly rely on URL properties. 
For example, shared information edges, that consider sharing of cookie values via query strings in the URL, are affected by URLs manipulations.

\subsection{Structure mutation attacks} \label{sec:attacks}

Next, we evaluate \name's robustness against structure mutations. 
We assume that the adversarial third-party has unrestricted black box access to the \name's classifier, i.e., the adversary can make unlimited queries and observe \name's classification output.
This access enables the adversary to validate the effect of their structure mutations.

\para{Attack details.} 
We assume that the adversary can mutate the structure of a web page through resource addition, re-routing, and obfuscation. 
Moreover, we assume that the adversary also performs content mutations, to maximize its chance of success
Resource addition entails addition of new resources, such as images and scripts. 
Resource re-routing entails re-organization of existing redirect chains, i.e., dispersing a redirect chain in a sequence of \texttt{XMLHttpRequest's} through one or multiple scripts. 
Resource obfuscation entails obfuscation of cookie or query string parameter values of existing resources, i.e., encoding or encrypting cookie or query string parameter values in a format that is not detected by \name's implementation, before sharing them in network requests. 
To remain stealthy, we assume that the adversary does not delete functional content from the web page that could damage usability.

It is important to note that even simple mutations, such as adding a single element to the web page, can significantly change graph properties and impact several features.
For example, the addition of a child node causes a cascading effect.
It increases the number of descendants of all the parent nodes in the branch, all the way up to the root node, and also impacts their centrality. 
Thus, the result of such simple mutations can become unpredictable and hard to control by the adversary:
It can cause unintended classification changes for nodes under and outside the control of the adversary.
Complex mutations, such as adding a combination of nodes at once, further complicate having control on the number of unintended classification changes.
In our evaluation, we only consider atomic mutations, i.e., addition, re-routing, or obfuscation of individual resources.

 \newcommand{\Goriginal}{G_0}
 \newcommand{\maxiter}{\texttt{max\_iter}}
 \newcommand{\feat}{\texttt{x}} 
 \newcommand{\pred}{\texttt{y}}
 \newcommand{\desired}{\texttt{d}}
 \newcommand{\undesired}{\texttt{u}}

\para{Mutation algorithm.} 
We capture the adversary's unrestricted black box access to classifier by implementing a greedy random algorithm to find suitable mutations.
This kind of algorithm is extensively used in the literature due to its simplicity and practicality~\cite{Zugner18AdversarialKDD, Hou19AlphaCyberIKM, Wang18FakeNodes}.
The algorithm (formally described in Appendix \ref{appx:mutation-algorithm}) iteratively mutates \name's graph representation.
At each step, it adds, re-routes, or obfuscates the resource that provides the best trade-off between \textit{desired} (\texttt{ATS} to \texttt{Non-ATS}) and \textit{undesired} (\texttt{NON-ATS} to \texttt{ATS}) classification switches.
Resource addition is simulated by adding nodes to a randomly selected leaf nodes in the graph. 
Resource re-routing is simulated by adding each request, in a redirect chain, as an individual node to one or more randomly selected scripts. 
Resource obfuscation is simulated by replacing stored values in URLs with an encoding that is not detected by \name.

\subsection{Empirical evaluation} 

\para{Experimental Setup.} 
\label{subsec:exp_setup}
To evaluate \name's robustness, we must rebuild the graph and recompute the features after each mutation. 
To keep the evaluation time reasonable, we sample \robgraphs web pages from our dataset, and we limit the graph growth to \robgrowth.
To ensure that this sample is representative of our dataset, we divide graphs into \robbins bins according to their size and sample \robsamples web pages from each bin. 
We only consider web pages that have 250 or fewer nodes (i.e., 80\% of the dataset; see Appendix~\ref{sec:size-distribution} for the full distribution).
We exclude large web pages to avoid exceptionally long evaluation times. 
For each web page, we designate the adversary as the third party with the highest number of resources classified as \texttt{ATS}.
It is noteworthy that the adversary with the highest number of \texttt{ATS} resources has an opportunity to do maximum damage. 

In this dataset, the median evaluation time per web page was 29.08 minutes, with 39\% of the pages taking more than an hour to run. 
Even though this is a simulation, the computational cost is directly proportional to the operational cost for the adversary.
The adversary must consume additional CPU cycles and memory and in the case of node addition, send additional network requests, thereby increasing the cost of their attack.

 \newcommand{\AdvT}{\texttt{ATS}_{\texttt{Adv}}}
 \newcommand{\AdvNT}{\texttt{Non-ATS}_{\texttt{Adv}}}
 \newcommand{\NAdvT}{\texttt{ATS}_{\texttt{Web}} }
 \newcommand{\NAdvNT}{\texttt{Non-ATS}_{\texttt{Web}}}

\para{Success metrics.}
To measure adversary's success, we define the following terms:

\begin{description}
    \item $\NAdvT$: Number of nodes classified as \texttt{ATS}.
    \item $\AdvT$: Number of adversary nodes classified as \texttt{ATS}.
    \item $\NAdvNT$: Number of nodes classified as \texttt{Non-ATS}.
    \item $\AdvNT$: Number of adversary nodes classified as \texttt{Non-ATS}.
    \item \texttt{desired}: Number of nodes switching from $\AdvT$ to $\AdvNT$.
    \item \texttt{undesired}: Number of nodes switching from $\NAdvNT$ to $\AdvT$.
    \item \texttt{neutral}: Number of nodes switching from \texttt{ATS} to \texttt{Non-ATS} for non-adversary nodes.
    \item \textit{Success rate}: Desired changes from the adversary's point of view. It is calculated as $\texttt{desired}/\AdvT$.
    \item \textit{Collateral damage}: Undesired changes from the adversary's point of view. It is calculated as $\texttt{undesired}/(\AdvNT+\NAdvNT)$.
    \item \textit{Other changes}: Non-consequential changes from the adversary's point of view. It is calculated as $\texttt{neutral}/\NAdvT$.
\end{description}

We illustrate the node switches, with the mutation algorithm, for an example graph in Appendix~\ref{subsec:mutations}.

\subsubsection{Adversary's success} 
\label{subsec:limadv}
\begin{figure*}[!htpb]
  \centering
  \subfigure[]
  {
      \includegraphics[width=0.3\textwidth]{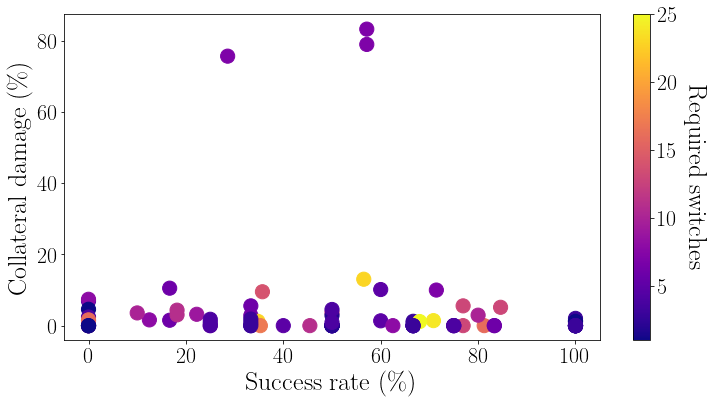}
              \label{fig:scatter_robustness}
  }
  \subfigure[]
  {
      \includegraphics[width=0.3\textwidth]{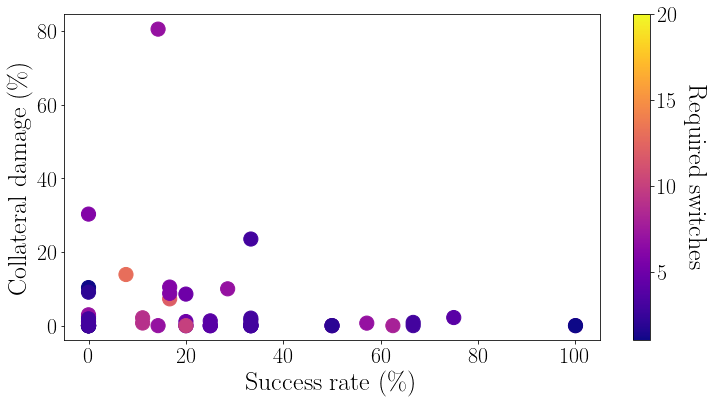}
              \label{fig:rec_redirects}
  }
  \subfigure[]
  {
      \includegraphics[width=0.3\textwidth]{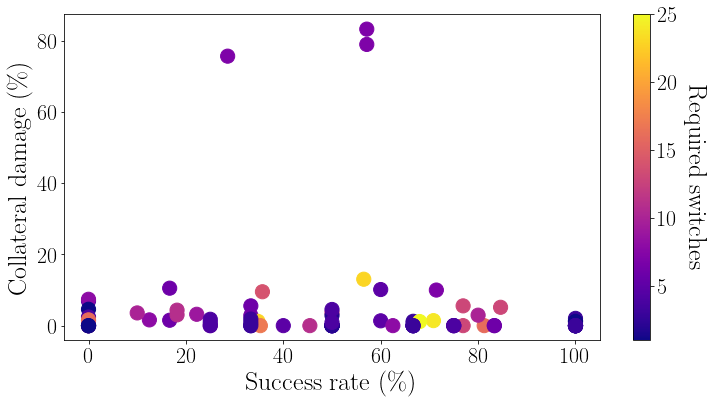}
      \label{fig:indirect_ancestors}
  }
\caption{Adversary's success rate vs. collateral damage for each web page in the test data at 20\% graph growth. Figure \ref{fig:scatter_robustness} represents all mutations, \ref{fig:rec_redirects} represents only structure mutations, and \ref{fig:indirect_ancestors} represents only resource re-routing and obfuscation mutations. Colored circles represent the number of required switches.}
\label{fig:scatter_robustness-all}
\vspace{-10pt}
\end{figure*}

We assume that the adversary neither colludes with other third parties nor with the first party and can only perform mutations on the nodes and edges it controls.
We conduct the attack on \robgraphs web pages.
We note that increasing the number of graph mutations increases the adversary's mean success rate from 38.6 $\pm$ 33.01 (median: 33.33) at 5\% graph growth to 52.48 $\pm$ 33.4 (median: 50.00) at 20\% graph growth.
The classification switches lead to a decrease in the overall classification accuracy by 1.5\%, recall by 8.85\%, and precision by 2.29\%.
However, the adversary's success comes at a cost of collateral damage.
The average collateral damage rises from 2.17 $\pm$ 11.19 to 3.88 $\pm$ 13.55 (median: 0).
In Figure~\ref{fig:scatter_robustness-all}, we illustrate the trade-off between success rate and collateral damage at 20\% graph growth.
The x-axis represents success rate, the y-axis represents collateral damage, and circles represent a trade-off between the two.
The circles' color represents $\AdvT$ or the number of classifications the adversary has to switch for the particular web page. The lighter the color, the more switches are required, i.e, the cost of success increases.
For the web pages in this dataset, the adversary has, on average, $\AdvT=$ \robreqmean $\pm$ \robreqstd nodes classified as \texttt{ATS}.
For certain pages, the $\AdvT$ can be as high as 26.

Ideally, the adversary wants to be at the bottom right of the graph, where it achieves 100\% success rate with zero collateral damage.
The adversary is able to reach its ideal target on only 13 web pages, which only required four switches. 
The adversary is able to achieve 50\% or more success on 61 of the tested web pages. 
Together, they amount to 240 nodes switched, with 45 of these pages having non-zero collateral damage.
On the other hand, we have 9 web pages that had a higher collateral damage than success rate: a net negative effect of the mutation.
Out of these, 6 web pages had 0\% success rate with non-zero collateral damage, and 3 web pages had a large collateral damage $>$ 75\% (with one web page hitting 83\%). 

Overall, our evaluation shows that even in the case of an unrealistic adversary that has the capabilities to manipulate structure features at will, and also the operational power to do so for a large number of iterations, there is no guarantee of perfect success.

\para{Breakage.} 
If undesired changes affect benign resources that are essential to the correct functioning of the web page, even a small collateral damage can break the page. 
This may have large impact on trackers.
If users leave the broken web pages, the adversary cannot track them or show them ads.
We define website breakage as degradation in usability of the website.
We say there is \texttt{major} breakage if the user is unable to complete the primary functionality of the web pages (e.g. login, search or page navigation). 
If the user is unable to complete a secondary functionality of the web pages (e.g. comment or review), we consider that there is \texttt{minor} breakage. 
Otherwise, we consider that the web page does not have any breakage. 
We quantify breakage on all of the 21 web pages where the adversary experiences undesired classification switches.\footnote{In total, the adversary experiences undesired classification switches on 45 web pages. However, 24 web pages no longer serve the switched \texttt{ATS} resources.}
We open these web pages side by side on stock Firefox and a Firefox configured with an extension that blocks the URLs that switched classification, and we compare them side by side to identify any visual signs of breakage. 

We ask two reviewers to perform the analysis.
Our reviewers attain an agreement of 90.46\% in their evaluation. 
They find that the undesired classification switches cause \texttt{major} breakage on 3 and \texttt{minor} breakage on 2 web pages.
This breakage mostly happens when the first-party resources are switched from \texttt{Non-ATS} $\rightarrow$ \texttt{ATS}.

\para{Careless adversary.} 
If the adversary is not concerned with changes to any non-adversarial nodes, their collateral damage decreases.
The adversary still does not want their own content to be blocked, so it will optimize against their own nodes switching to \texttt{ATS}.
This change in strategy updates the collateral damage calculation to:  $\texttt{undesired}/\AdvNT$.

As per our modified definition, the web pages on which all of the adversary nodes are classified as \texttt{ATS}, there can be no collateral damage; we note 55 such web pages.
For the remaining 45 web pages, where the adversary can experience collateral damage, the mean growth in success rate does not change much from the previous scenario, but naturally the trade-off is better.
Further, out of 45 web pages, only 8 web pages have collateral damage, as compared to 27 web pages that had collateral damage as per our original definition.
Out of these 8 web pages, 4 had a higher collateral damage than success rate (net negative effect), and 6 web pages have a large collateral damage > 20\% (with 2 web pages hitting 100\%).
Thus, even when an adversary is not concerned about collateral damage to other parties they are not significantly more successful in subverting \name.

\para{Collusion with the first party.}
So far, we have assumed that the adversary is a single third party that does not collude with other third parties or the first party.
If we assume the adversary colludes with both, the adversary can add child nodes to \emph{any} node in the graph.
This is a much stronger adversary than in Section~\ref{subsec:limadv}, where in each iteration the adversary can only test a random subset of the options.
Realistically, such a powerful collusion would be difficult to implement, as it would require coordination and cooperation among multiple parties to ensure that the mutation is feasible.

We repeat our experiment, but we now allow the adversary to consider all possible mutation options on any node, and pick the best one in each iteration.
These experiments take longer to run (see Appendix~\ref{sec:run-times}), so we only analyze \supadvgraphs web pages whose graphs have at most 50 nodes. 
We see that collusion enables the adversary to have a slightly higher success rate (63 pages with success rate > 50\% as compared to 60 for the non-colluding adversary) and lower collateral damage (9 pages with damage >0\% compared to 18 pages for a non-colluding adversary).
These results are described in detail in Appendix~\ref{sec:robustness-additional}).

\subsubsection{Impact of mutation choice} \label{subsec:mutationchoice}
Next, we evaluate the adversary's preference in selecting the most useful mutations. 
We notice that the adversary picks resource addition 81.70\%, resource re-routing 17.26\%, and resource obfuscation 0.04\% of the time. 
Resource obfuscation is rarely chosen by the adversary because the graph already has content manipulations applied, and these manipulations have already severed many of the edges that would be severed by resource obfuscation. 
To separate out the impact of different mutations, we conduct two additional experiments: (1) where the adversary can only perform resource addition, and (2) where the adversary can only perform resource re-routing and obfuscation.

We exclude 33 of the web pages for experiment 2 because these web pages do not have re-route-able or obfuscate-able resources. 
For the remaining 67 pages, we see that the re-routing/obfuscation mutations (Figure~\ref{fig:indirect_ancestors}) are more effective than addition mutations (Figure~\ref{fig:rec_redirects}).
Re-routing/obfuscation not only yields higher success rates for the adversary, but also results in lower collateral damage.
This is unsurprising because these mutations target information sharing patterns which are distinctive of trackers; changing these patterns removes an important signal for the classifier (see Table~\ref{tab:importance_webgraph_nodomain}).

However, in practice, resource re-routing and obfuscation would entail high costs for the adversary since they involve the manipulation of identifier sharing patterns. 
Specifically, the adversary would have to coordinate with other parties on changes to these patterns, and redesign how they perform tracking in order to perform these mutations.
The success of these mutations also depends on the degree to which flows are captured by the instrumentation used to create the graph.
\name's instrumentation approximates information flows and will not capture all attempts by an adversary to use re-routing and obfuscation.
We argue that this is not a fundamental flaw in \name's architecture but a limitation in our implementation that approximates information flow (Section \ref{subsec:infoflow}). 
A fully fledged instrumentation would make these manipulations much more difficult to deploy.
See Section~\ref{sec:improvements} for an extended discussion.
Resource addition has fewer costs for the adversary since it does not involve coordination with additional parties.
This manipulation is not affected by the type of implementation because it is not related to the flow of identifiers.

\subsubsection{Comparison with \adgraph.} 
\label{subsec:adgraphrob}
We also evaluate whether \name, in addition to having superior classification performance, offers robustness benefits over \adgraph.
For this comparison, we only use \adgraph's structural features, as we already demonstrated that content features are not robust.
Because \adgraph does not have features based on flow information, we only perform resource addition.
We find that the adversary has greater success against \adgraph than \name, but also suffers from more collateral damage (Figures~\ref{fig:adgraphstruct} and \ref{fig:webgraphstruct} in Appendix~\ref{sec:robustness-additional}).
This is because the structural effects of node additions are hard to control, as explained in Section~\ref{sec:attacks}.
Since the former is beneficial to the adversary but the latter is not, it is not clear-cut as to whether one system provides more robustness than the other.
In summary, our results indicate that mutations to the structure of the graph are harder for an adversary to control than content mutations.
It is not trivial for an adversary to produce the desired classification switches for their resources without producing any undesired changes. 
This makes \name, without content features, more robust to adversarial evasion attacks than prior approaches.
Within the structural mutations, re-routing/obfuscating resources target information flow and are a more effective strategy than adding resources.
At the same time, performing these mutations is not trivial for the adversary since they involve coordination with multiple parties.

\section{Limitations}
\label{sec:improvements}
In this section, we discuss limitations of \name's design, implementation, and evaluation.

\para{Completeness.} 
For efficiency reasons, \name focuses on a limited subset of the browser's API surface, such as HTTP cookie headers, \texttt{document.cookie}, and \texttt{window.localStorage}.
\name's implementation is also geared towards capturing client-side information that is pertinent to stateful tracking. 
However, techniques used by ATS need not to be limited to these APIs or to stateful tracking.
Some ATS have started to use stateless tracking techniques, such as browser fingerprinting, which use APIs that are not currently covered by our instrumentation \cite{Iqbal21FingerprintingSP,Das18SmartphontCCS,MavroudisHFMKV17}. 
To account for these techniques, \name's instrumentation must be extended to include the corresponding APIs.

\name's manually designed graph representation and feature set capture the most well-known information sharing patterns. 
The limits of these approach are shown in Section~\ref{subsec:mutationchoice}, where we show that an adversary capable of hiding or obfuscating traditional sharing flows has a better chance to bypass \name than doing structure modifications.
This limitation is, however, linked to our implementation choices.
To increase \name's coverage of sharing behaviors, if suffices with increase the instrumentation to cover more information flows.
Ideally, we would instrument full-blown information flow tracking. 
Such expansion would incur prohibitive runtime overheads (up to 100X-1000X~\cite{Hedin14JSFlowSAC}) and its complexity makes it hard to integrate in the browser~\cite{Chudnov2015FlowCCS, Chen18MystiqueCCS, Stock14XSSProtectionUsenix, Lekies13FlowXSSCCS}. 
Nevertheless, the design of \name permits that the instrumentation to be upgraded gradually, as ATS evolve in response to our evasion protection techniques, increasing the cost of evasion without fundamentally changing the detection approach. 

\para{Robustness analysis.}
Inspired by previous work on graph-based detection evasion~\cite{Zugner18AdversarialKDD, Hou19AlphaCyberIKM, Wang18FakeNodes}, we use a greedy algorithm to attack \name.
This algorithm only considers the best mutation in each iteration, and not the best overall mutation.
Thus, it is not guaranteed to find the optimal mutation sequence that would lead to the best adversary performance.
We note however that, as our experiments on small websites show, even exhaustive search does not lead to perfect success.
We expect adversaries to try alternative algorithms to improve their success rates.
However, any alternative that is close to exhaustive search will become prohibitively expensive for the adversary when the web page graph is large.

Another option for the adversary would be to perform more sophisticated graph mutations instead of the simple node additions that we perform. 
An adversary could tailor their mutations to the page's graph structure by studying how their node changes affect the graph properties of the web page. 
However, this requires that the rest of the graph (i.e., the portions outside of the adversary's control) remaining unchanged. 
Realistically, it would be difficult for an adversary to coordinate with other parties to generate these changes.

Finally, we note that the dynamism of modern websites \cite{Butkiewic11IMCwebsitecomplexity} complicates the process for the adversary.
Web pages change often, sometimes on every load.
Even if the adversary manages to find an appropriate set of mutations, those mutations may be invalid the next time the page is reloaded.

\section{Conclusion}
\label{sec:conclusion}

In this paper, we showed that state-of-the-art ad and tracker blocking approaches are susceptible to evasion due to their reliance on easy-to-manipulate content features.
We then showed that information sharing patterns in online advertising and tracking can instead be leveraged for robust blocking. 
Specifically, our proposed \name builds a cross-layer graph representation to capture such information flows and train a machine learning classifier for accurate and robust ad and tracker blocking.
Our results showed that it is non-trivial to evade \name's classifier without causing unavoidable collateral damage.

While it is not foolproof, we believe that \name raises the bar for advertisers and trackers attempting to evade detection.  
We foresee that advertising and tracking services would need to significantly re-architecture their information sharing patterns to achieve long-lasting evasion against \name. 
We note, however, that introducing new information flows may be quite complicated, as they may require collaboration among the first-party and numerous third-parties on a typical web page.

\bibliographystyle{unsrt}
\bibliography{bibliography}

\appendix
\section{Comparison between \adgraph and \name features} \label{sec:adgraph-comparison}

Table~\ref{tab:dataflow_features} compares and contrasts the features used in \name and \adgraph.
 \name does not use content features. 
 Graph size, Degree and Centrality features come under both structure and flow categories, since they include graph properties that are based on both normal (structure feature) and shared information edges (flow feature).
 \name uses both types of edges, whereas \adgraph uses only normal edges.
  Some structural features used in \adgraph are not used in \name due to \name being adapted for offline use, whereas the features are useful in an online context. 

\begin{table*}[h!]
  \centering
  \caption{\name features comparison with \adgraph. \fullcirc\xspace indicates that a feature is present. \name calculates Graph size, Degree and Centrality features using both normal and shared information edges. The former comes under structural features while the latter comes under flow features.}
  
  \resizebox*{2\columnwidth}{!}{%
  \begin{tabular}[c]{l|l|c|c}
   \toprule
     \textbf{Feature} & \textbf{Type} & \textbf{\name} & \textbf{\adgraph}\\
    \midrule
        Request type (e.g. iframe, image) & Content &  & \fullcirc\\
    Ad keywords in request (e.g. banner, sponsor) & Content &  & \fullcirc \\
    Ad or screen dimensions in URL & Content &  & \fullcirc \\
    Valid query string parameters & Content &  & \fullcirc \\
    Length of URL & Content &  & \fullcirc \\
    Domain party & Content &  & \fullcirc \\
    Sub-domain check & Content &  & \fullcirc \\
    Base domain in query string & Content &  & \fullcirc \\
    Semi-colon in query string & Content &  & \fullcirc\\
    \midrule
       Graph size (\# of nodes, \# of edges, and nodes/edge ratio) & Structure & \fullcirc & \fullcirc \\
    Degree (in, out, in+out, and average degree connectivity) & Structure & \fullcirc & \fullcirc \\
       Centrality (closeness centrality, eccentricity) & Structure & \fullcirc & \\
     Number of siblings (node and parents) & Structure &  & \fullcirc \\
    Modifications by scripts (node and parents) & Structure &  & \fullcirc \\
    Parent’s attributes & Structure &  & \fullcirc \\
    Parent degree (in, out, in+out, and average degree connectivity) & Structure &  & \fullcirc\\
    Sibling’s attributes & Structure &  & \fullcirc\\
    Ascendant’s attributes & Structure & \fullcirc & \fullcirc\\
    Descendant of a script & Structure & \fullcirc & \fullcirc\\
    Ascendant’s script properties & Structure & \fullcirc & \fullcirc \\
    Parent is an eval script & Structure & \fullcirc & \fullcirc \\
    \midrule
    Local storage access (\# of sets, \# of gets) & Flow (storage) & \fullcirc &  \\
    Cookie access (\# of sets, \# of gets)  & Flow (storage) & \fullcirc & \\
   Requests (sent, received)  & Flow (network) & \fullcirc & \\
   Redirects (sent, received, depth in chain)   &  Flow (network) & \fullcirc & \\
   Common access to the same storage node & Flow (shared information) & \fullcirc & \\
   Sharing of a storage node's value in a URL & Flow (shared information) & \fullcirc & \\
     Graph size (\# of nodes, \# of edges, and nodes/edge ratio) & Flow (shared information) & \fullcirc &  \\
    Degree (in, out, in+out, and average degree connectivity) & Flow (shared information) & \fullcirc &  \\
    Centrality (closeness centrality, eccentricity) & Flow (shared information) & \fullcirc & \\
   \bottomrule
     \end{tabular}
     }
  \label{tab:dataflow_features}
\end{table*}

\section{Distribution of graph sizes }
\label{sec:size-distribution}

\begin{figure}[H]
\centering
\includegraphics[width=0.48\textwidth]{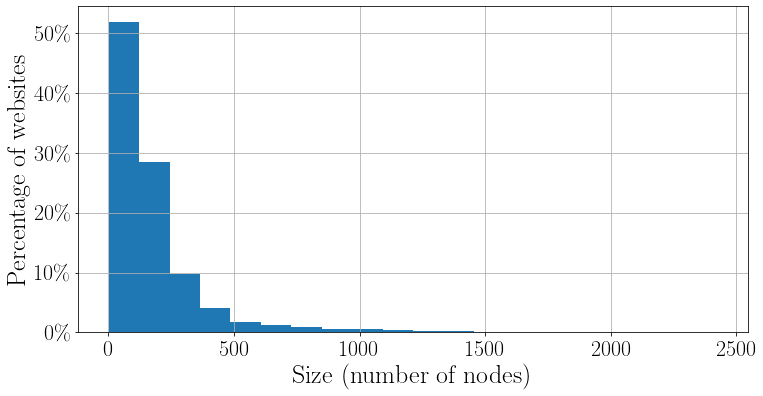}
\caption{Distribution of number of nodes in the graph representations of the web pages in the dataset. 80\% of the web pages have 250 nodes or fewer.}
\label{fig:sizes}
\end{figure}

Figure~\ref{fig:sizes} shows the distribution of number of nodes in the graph representations of the web pages in our dataset. Since 80\% of web pages have 250 nodes or fewer, we sample from this subset in our structural mutation experiments in Section~\ref{sec:robustness}.

\section{Experimental run times}
\label{sec:run-times}

\begin{figure}[H]
\centering
\includegraphics[width=0.48\textwidth]{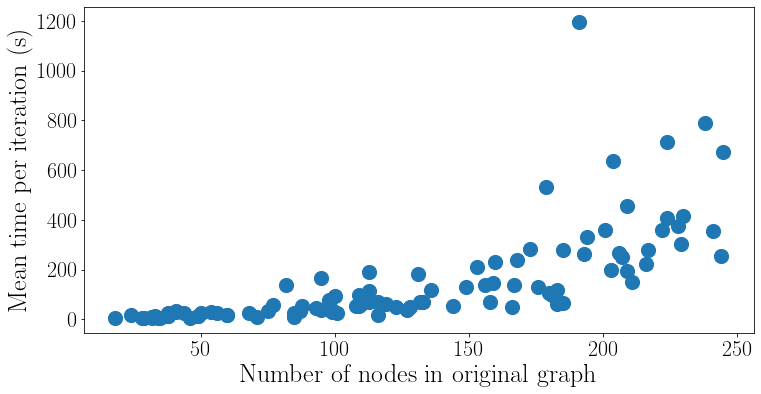}
\caption{Mean time per iteration vs graph size for the experiment without collusion. Standard deviation over all the iterations for each graph was less than 2\%. Larger graph sizes take longer time for each iteration.}
\label{fig:periter}
\end{figure}

\begin{figure}[H]
\centering
\includegraphics[width=0.48\textwidth]{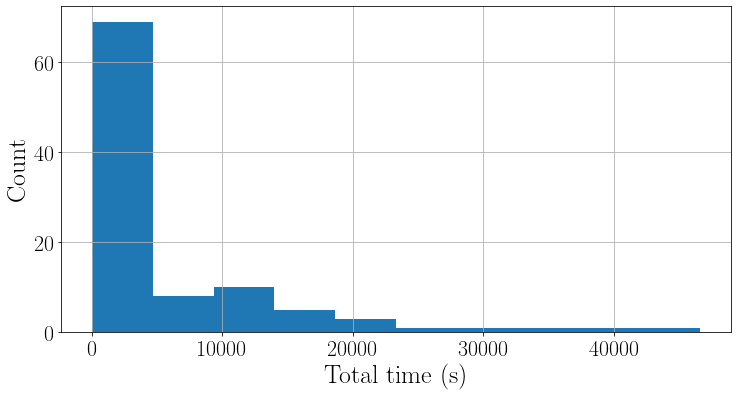}
\caption{Distribution of total run time for the experiments in Section~\ref{subsec:limadv}. 39\% of the graphs in our dataset have a run time of more than an hour.}
\label{fig:time_limadv}
\end{figure}

Figures~\ref{fig:periter} and ~\ref{fig:time_limadv} describe the run times for the experiment described in Section~\ref{subsec:limadv} (adversary without collusion). 
Figure~\ref{fig:periter} shows the impact of graph size on each iteration of the experiment. 
As expected, smaller graphs have lower run times since features have to be calculated over a smaller number of nodes.
Note that graph size is not the only contributing factor to run times. 
Other factors such as the complexity of the structure and flow behaviors would also contribute towards time spent in each iteration, which is why we observe variations in iteration time among graphs of the same size.
We see that the mean time per iteration can be as high as $\approx$ 1200 seconds (median is $\approx$ 68 seconds).
Figure~\ref{fig:time_limadv} shows the total experiment time over all iterations for a graph.
Since we increase the sizes of graphs by 20\% of their original size, bigger graphs will have a larger number of iterations.
In our dataset, the maximum time taken for an experiment is 46654.19 seconds, the minimum is 15.67 seconds, and the median is 1745.11 seconds. 
39\% of the graphs in our dataset have a run time of more than an hour.

\begin{figure}[H]
\centering
\includegraphics[width=0.48\textwidth]{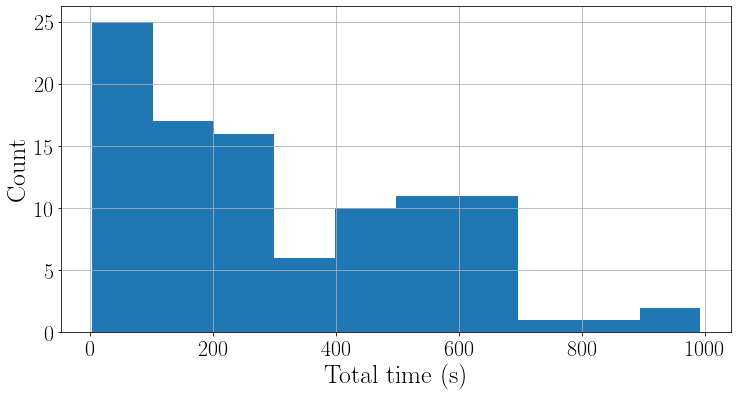}
\caption{Distribution of total run time for the experiments in Section~\ref{subsec:limadv}. The experiment time can be as high as 996.67 seconds for a graph size  $<=$ 50 nodes.}
\label{fig:time_supadv}
\end{figure}

Figure~\ref{fig:time_supadv} shows the total experiment time over all iterations for the experiment described in Section~\ref{subsec:limadv} (collusion with first party). 
The median time is 265.03 seconds, with the maximum time going up to 992.67 seconds, despite the maximum graph size being only 50 nodes.
In comparison, for the adversary without collusion, for graph sizes up to 50 nodes, the median is 21.46 seconds and the maximum is 221.51 seconds.
Since the adversary considers all nodes in the graph as potential parents, each iteration takes a longer amount of time.

\section{Graph Mutation algorithm}
\label{appx:mutation-algorithm}

In each iteration, the algorithm mutates \name's graph representation and probes the model for classification decisions. 
The algorithm takes the following inputs:
a graph representation of a web page, $\Goriginal$, consisting of all the nodes in the graph; 
a set of nodes and edges $T$ of size $l_{T}$, representing the resources loaded by the adversary, hereafter referred to as the adversary resources; 
a trained classifier $M$ that identifies ATS in \name;
and a maximum number of iterations that the algorithm can run, $\maxiter$.

\begin{algorithm}[t]
\begin{algorithmic}[1]
 \renewcommand{\algorithmicrequire}{\textbf{Input:}}
 \renewcommand{\algorithmicensure}{\textbf{Output:}}
 \REQUIRE $\Goriginal,T,C,M,\maxiter$
 \FOR{v $\in \Goriginal$}
 \STATE $\feat_{\Goriginal} \gets \texttt{ExtractFeatures}(v) \  \forall \  v  \in  \Goriginal$
  \STATE $\pred_{\Goriginal} \gets \texttt{Classify}(M,\feat) \ \forall \ x \ in \  \feat_{\Goriginal}$
  \ENDFOR
  \STATE $G \gets \Goriginal$ 
  \STATE $i \gets 0$
  \STATE $\texttt{graph-info} = []$
  \WHILE {$i < \maxiter$}
  \FOR{$t \in T$}
  \STATE $G_t \gets \texttt{MutateGraph}(G, t)$
  \STATE $\feat_t \gets \texttt{ExtractFeatures}(v) \  \forall \  v \in G_t$
  \STATE $\pred_t \gets \texttt{Classify}(M,\feat)  \  \forall \  x \  in \  \feat_t$
  \STATE $\desired, \undesired\ \gets \texttt{GetDesiredAndUndesired}(\pred_t, \pred_{\Goriginal}$)
  \STATE $\Delta_t = \desired - \undesired$
  \STATE $\texttt{graph-info}[t] \gets (\Delta_t,t,G_{t})$
 \ENDFOR
 \STATE $G \gets G_{t}$ in \texttt{graph-info}[t] with largest $\Delta_t$
 \STATE $T \gets  \texttt{UpdateAdv}(T, t \in \texttt{graph-info}[t])$ 
 \STATE $T \gets \texttt{sample}(T, l_{T})$
 \STATE $i \gets i + 1$
  \ENDWHILE
 \end{algorithmic} 
 \label{algo:mutation}
 \caption{Greedy random graph mutation. $\Goriginal$ is a web page representation, $T$ is the set of $l_{T}$ nodes and edges controlled by the adversary, $M$ is a trained model, and $\maxiter$ is the maximum number of operations.}
\end{algorithm}

The algorithm processes the input as follows:
It first uses the classifier $M$ to obtain classifications of all nodes in the original graph $\Goriginal$ (lines 1--4 in Algorithm~\ref{algo:mutation}).
Second, it iterates over the steps from lines 9--20 $\maxiter$ times.
In each iteration, every adversary node tries resource addition, resource re-routing, and obfuscation, and produces a new mutated graph, $G_{i}$ (line 11). 
%
%
%
Third, it extracts features from the mutated graph $G_{i}$ and uses them to classify all the nodes in this graph (lines 11--12). 
Fourth, it compares the predictions in the original and mutated graphs to obtain the number of desired and undesired switches (line 13).
We assume an adversarial goal for which \emph{desired} switches are all those in which an adversary node is switched from \texttt{ATS} to \texttt{Non-ATS}, whereas \emph{undesired} switches are all those where any \texttt{Non-ATS} node is switched to \texttt{ATS} node. 
We call the total number of adversarial \texttt{ATS} nodes whose prediction the adversary wishes to change to \texttt{Non-ATS} the number of \emph{required} switches.
The switching of nodes not under the adversary's control from \texttt{ATS} to \texttt{Non-ATS} do not affect the adversary.
These switches are, therefore, neither desired nor undesired.
Finally, the adversary chooses the mutation that provides the best result, i.e., the one with the best trade-off between desired and undesired switches (lines 14--15). 
The adversary updates its $T$ based on the chosen mutation (line 18). 
To keep memory and run time manageable, at the end of every iteration the algorithm randomly samples $l_{T}$ adversarial nodes and edges from $T$ (line 19) to be considered in the next iteration.

\begin{figure}[t!]
\centering
\includegraphics[width=0.48\textwidth]{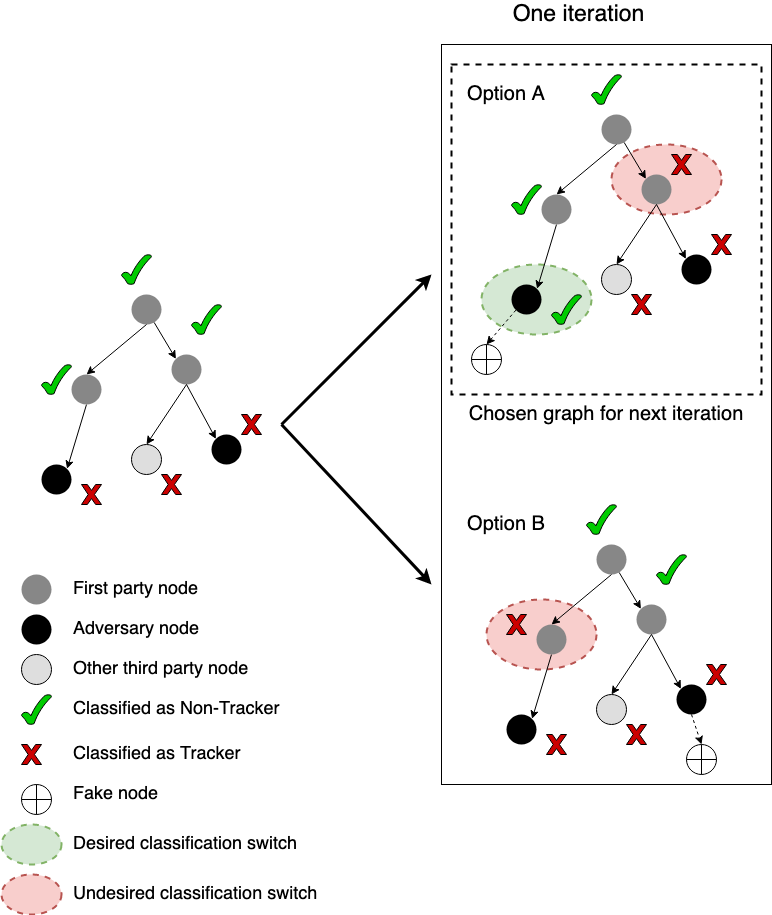}
\caption{One iteration of the greedy random mutation algorithm. In this iteration the algorithm selected two adversarial nodes to add a child (Option A: bottom left node, top; Option B: bottom right node, top;). In option A, the change leads to one desired (green circle) and one undesired (red circle) modification. In option B, the change only causes one undesired modification. The algorithm would pick the graph with best trade-off between desired and undesired switches for the next iteration: i.e., Option A.}
\label{fig:mutation}
\end{figure}

\textbf{Mutation example.}
An example iteration of the greedy random algorithm, using resource addition as the mutation, is shown in Figure~\ref{fig:mutation}.
The algorithm proposes two mutations (right) to the initial graph (left), and chooses the mutation that provides the best trade-off between the desired and undesired classification switches. 
Figure~\ref{fig:mutation} also illustrates that even a simple node addition may lead to unintended changes in classification decisions: the adversary may change the classification decisions for some nodes from \texttt{ATS} to \texttt{Non-ATS}, but as a side effect nodes previously classified as \texttt{Non-ATS} are now classified as \texttt{ATS}.
%

\section{Mutations on a single web page.} \label{subsec:mutations}
\begin{figure}[t!]
\centering
\includegraphics[width=0.48\textwidth]{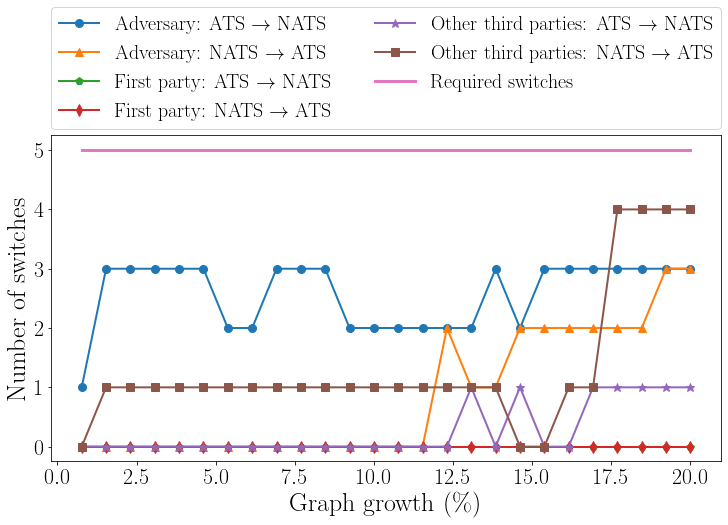}
\caption{\looseness=-1 Example breakdown of classification switches for the adversary's and other nodes on the graph. \texttt{NATS} is shorthand for \texttt{Non-ATS}. $\AdvT$ = 5 (pink line), $\AdvNT$ = 7, $\NAdvT$ = 62, $\NAdvT$ = 13 (not shown in plot). At 20\% growth, the adversary achieves 3 \texttt{desired} switches, 7 \texttt{undesired} switches and 1 \texttt{neutral} switch. This leads to a success rate of 60\%, a collateral damage of 10.14\% and other changes of 7.7\%.}
\label{fig:breakdown}
\end{figure}

To illustrate how mutations result in classification switches, we take as an example a web page in which the third party with the highest number of \texttt{ATS} resources is \url{assets.wogaa.sg}, which has 12 nodes in the graph.
%
Figure~\ref{fig:breakdown} shows the breakdown of classification switches as the adversary mutates the graph using the greedy mutation algorithm.
The $\AdvT$ or the number of classifications the adversary wants to switch is 5 (pink line \protect\tikz[baseline]{\protect\draw[line width=0.3mm, color=pink] (0,.6ex)--++(0.5,0) ;}).
From the adversary's point of view, adversarial nodes switching from \texttt{ATS} $\rightarrow$ \texttt{Non-ATS} are desired (blue line \protect\tikz[baseline]{\protect\draw[line width=0.3mm, color=blue] (0,.6ex)--++(0.5,0) ;\node[mark size=2pt,color=blue] at (0.25,0.1) {\pgfuseplotmark{*}};}), 
whereas adversarial nodes switching from \texttt{Non-ATS} $\rightarrow$ \texttt{ATS} are undesired (orange line \protect\tikz[baseline]{\protect\draw[line width=0.3mm, color=orange] (0,.6ex)--++(0.5,0) ;\node[mark size=2pt,color=orange] at (0.25,0.1) {\pgfuseplotmark{triangle*}};}).
We consider \texttt{Non-ATS} $\rightarrow$ \texttt{ATS} changes on non-adversarial nodes to be undesired because they may have unintended impact on the web page (red line \protect\tikz[baseline]{\protect\draw[line width=0.3mm, color=red] (0,.6ex)--++(0.5,0) ;\node[mark size=2pt,color=red] at (0.25,0.1) {\pgfuseplotmark{diamond*}};} and brown line \protect\tikz[baseline]{\protect\draw[line width=0.3mm, color=brown] (0,.6ex)--++(0.5,0) ;\node[mark size=2pt,color=brown] at (0.25,0.1) {\pgfuseplotmark{square*}};}).
For instance, a first party \texttt{Non-ATS} $\rightarrow$ \texttt{ATS} switch may break the web page. 
We note that, if the adversary's goal is to just create a denial of service and force the user to disable ad and tracker blocking, the adversary might be unconcerned about breakage. 
In our experiments, switches that do not affect the adversary, such as \texttt{ATS} $\rightarrow$ \texttt{Non-ATS} for non-adversary nodes, are neither considered desired nor undesired (purple line \protect\tikz[baseline]{\protect\draw[line width=0.3mm, color=purple] (0,.6ex)--++(0.5,0) ;\node[mark size=2pt,color=purple] at (0.25,0.1) {\pgfuseplotmark{star}};} and green line \protect\tikz[baseline]{\protect\draw[line width=0.3mm, color=green] (0,.6ex)--++(0.5,0) ;\node[mark size=2pt,color=green] at (0.25,0.1) {\pgfuseplotmark{pentagon*}};}).
There are two points worth highlighting from Figure~\ref{fig:breakdown}: 
(1) Even if an adversary achieves the maximum number of desired switches, the mutations may produce undesirable changes, to both the adversary's nodes and others. For instance, at 20\% of growth, 3 of the adversary's \texttt{ATS} nodes are classified as \texttt{Non-ATS}, but also 7 \texttt{Non-ATS} nodes (3 adversary and 4 non-adversary) switch to the undesired \texttt{ATS} classification.
(2) The evolution of desired and undesired switches is not monotonic, i.e. the classification of nodes may change in both directions as the adversary mutates the graph, resulting in increasing or decreasing counts. 
This finding reinforces our argument that it can be cumbersome for an adversary to create targeted structural mutations without any unintended consequences. 
Not only it is hard to predict how mutations will affect adversary's own desired classification, but also how those mutations may result in undesirable changes to others.

\section{\name robustness experiments} \label{sec:robustness-additional}

\begin{figure}[H]
\centering
\includegraphics[width=0.48\textwidth]{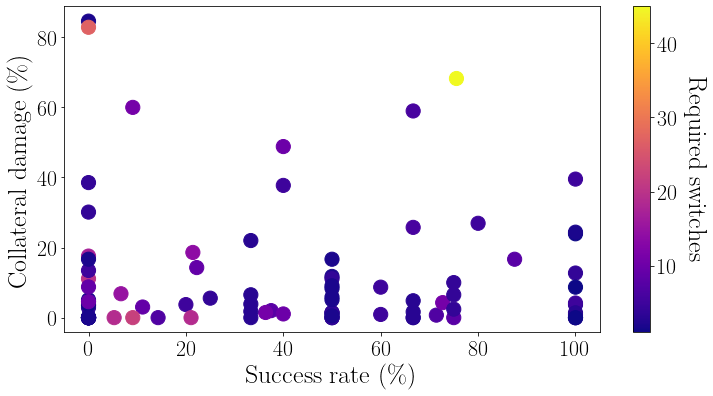}
\caption{Adversary's success rate vs. collateral damage for each web page in the test data at 20\% graph growth, for resource addition against \name. Color denotes the number of required switches.}
\label{fig:adgraphstruct}
\end{figure}

\begin{figure}[H]
\centering
\includegraphics[width=0.48\textwidth]{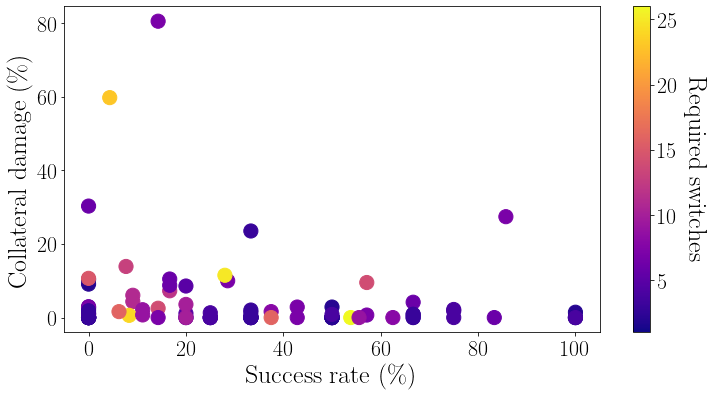}
\caption{Adversary's success rate vs. collateral damage for each web page in the test data at up to 20\% graph growth, for resource addition against \name. Color denotes the number of required switches.}
\label{fig:webgraphstruct}
\end{figure}

We show the success rate vs. collateral damage plots for the experiments described in Sections~\ref{subsec:limadv} and ~\ref{subsec:adgraphrob}.
Figures~\ref{fig:adgraphstruct} and ~\ref{fig:webgraphstruct} show the results for an adversary that performs only resource addition against \adgraph (with only structural features) and \name respectively.
\adgraph shows a higher number of successes for the adversary (44 pages with success rate > 50\% as compared to 30 for \name).
At the same time, \adgraph also shows a higher amount of collateral damage (which is not beneficial for the adversary) -- 66 pages with non-zero collateral damage, as compared to 47 for \name.
Hence, there is no clear-cut winner between the two classifiers in terms of robustness.
However, we do see that \adgraph has lower successes and higher collateral damage than \name against the powerful adversary that can do all mutations as shown in Figure~\ref{fig:scatter_robustness} (note that this adversary cannot be used against \adgraph since \adgraph does not use information flow edges), since this adversary targets the effective, but costly, information sharing patterns.

\begin{figure}[H]
\centering
\includegraphics[width=0.48\textwidth]{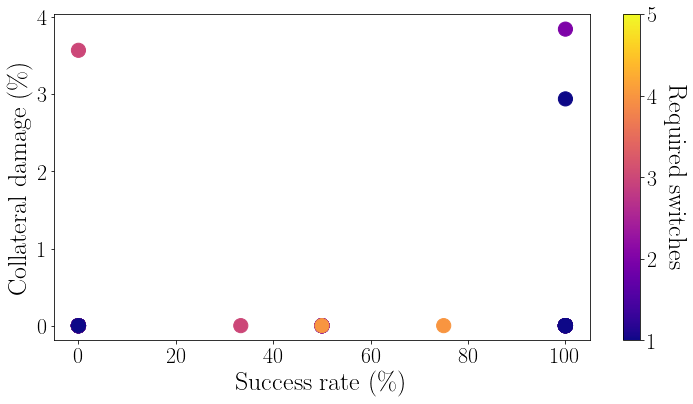}
\caption{Adversary's success rate vs. collateral damage for each web page in the test data at 20\% graph growth, for a colluding adversary. Color denotes the number of required switches.}
\label{fig:supadv}
\end{figure}

\begin{figure}[H]
\centering
\includegraphics[width=0.48\textwidth]{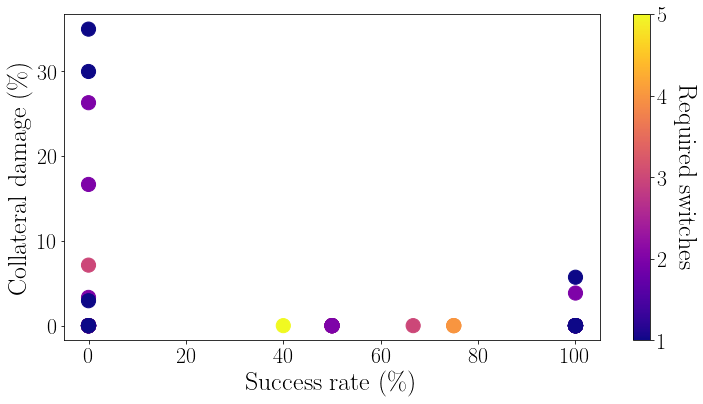}
\caption{Adversary's success rate vs. collateral damage for each web page in the test data at up to 20\% graph growth, for a non-colluding adversary. Color denotes the number of required switches.}
\label{fig:limadv}
\end{figure}

Figures~\ref{fig:supadv} and ~\ref{fig:limadv} show the results for an adversary that colludes against an adversary with no collusion (Section~\ref{subsec:limadv}).
A colluding adversary shows a higher number of successes (63 pages with success rate > 50\% as compared to 60 for the non-colluding adversary), and a lower collateral damage (9 pages with damage >0\% compared to 18 pages for a non-colluding adversary).

\clearpage
\onecolumn

\end{document}